\documentstyle[aps,epsfig,floats,amssymb]{revtex}
\draft

\begin{document}
\twocolumn[\hsize\textwidth\columnwidth\hsize\csname
@twocolumnfalse\endcsname

\title{Invariants and nonlinear fields in Spinor Gravity}

\author{C. Wetterich}

\address{
Institut f{\"u}r Theoretische Physik,
Philosophenweg 16, 69120 Heidelberg, Germany}

\maketitle

\begin{abstract}
Spinor gravity is a functional integral formulation of gravity based
only on fundamental spinor fields. The vielbein and metric arise as composite objects.
Due to the lack of local Lorentz-symmetry new invariants in the effective gravitational action lead to a modification of Einstein's equations. We discuss different geometrical viewpoints of spinor gravity with particular emphasis on nonlinear fields. The effective gravitational field equations arise as solutions to the lowest order Schwinger-Dyson equation for spinor gravity.
\end{abstract}
\pacs{PACS numbers: 12.10.-g; 04.20.Cv; 11.10.Kk  \hfill HD-THEP-04-05}

 ]

\section{Introduction}
\label{introduction}
The construction of a well defined functional integral for the metric is a notorious problem. In particular, such a non-perturbative definition of quantum gravity needs to implement properly the symmetry of general coordinate transformations (diffeomorphisms). As a possible way out one may use the observation \cite{Ak,Av,DS} that the metric and the vielbein could arise as composite objects in a theory with fundamental spinors. Since spinors transform as scalars under general coordinate transformations one may hope that the construction of a functional measure becomes easier as compared to fundamental metric degrees of freedom. Furthermore, the prospect of a possible unified theory which contains only fermionic degrees of freedom is quite intriguing. Not only the metric but also the gauge bosons and the Higgs scalar could arise as composite fields.

The requirement of a local polynomial spinor action which is invariant under general
coordinate transformations leads to the proposal \cite{HW1,SG1}
of ``spinor gravity''. In the
present paper we elaborate on the geometrical and symmetry aspects of this proposal\footnote{A large part of the material of the present paper was originally contained in the first version of ref. \cite{SG1}. The choice of a separate publication is motivated by a clear presentation of different aspects of spinor gravity.}. The construction of a local diffeomorphism invariant action is straightforward \cite{Ak,Av,DS}. As an example we may consider 
\begin{eqnarray}\label{AA0}
S_E&\sim&\int d^dx\det\big(\tilde{E}^m_\mu(x)\big)~,\nonumber\\
\tilde{E}^m_\mu&=&i\bar{\psi}\gamma^m\partial_\mu\psi.
\end{eqnarray}
Here $\psi(x)$ denotes Grassmann variables in the spinor
representation of the $d$-dimensional Lorentz group and we have introduced the associated
Dirac matrices $\gamma^m$ such that $\tilde{E}^m_\mu$ is a vector with respect to global
Lorentz rotations. Due to the derivative the bilinear $\tilde{E}^m_\mu$ transforms as a vector with respect to general coordinate transformations. The only possible choice for a diffeomorphism invariant action contracts $d$ derivatives with the totally antisymmetric tensor $\epsilon^{\mu_1\dots\mu_d}$.
Invariance under global Lorentz  rotations can be achieved similarly by contraction with
$\epsilon_{m_1\dots m_d}$. 

The action $S_E$ is invariant under {\em global} but not {\em local} Lorentz
transformations. This is an important difference as compared to the standard formulation
of gravity (``Einstein gravity'').
We will explore here the conceptual aspects of this difference while the phenomenological implications are discussed in detail in \cite{SG1}.
Actually, the action (\ref{AA0}) is not the only invariant
with respect to diffeomorphism and global Lorentz symmetry - other invariants are discussed below. In particular, invariants exhibiting also local Lorentz symmetry have been found recently \cite{CWST} and involve typically a high power of spinor fields. In the present paper we rather investigate the alternative possibility, namely that gravity
indeed exhibits only global and not local Lorentz symmetry. This will lead to a generalized
version of gravity. 

Within ``spinor gravity'' the
``global vielbein'' $E^m_\mu(x)$ can be associated to the expectation
value of the fermion bilinear $\tilde{E}^m_\mu(x)$. As usual
the metric obtains then by contraction
with the invariant tensor $\eta_{mn}$ which lowers the Lorentz indices
\begin{equation}\label{AA1}
E^m_\mu(x)=m\langle\tilde{E}^m_\mu(x)\rangle~,~g_{\mu\nu}(x)=E^m_\mu(x)E_{\nu m}(x).
\end{equation}
Here the arbitrary mass scale $m$ has been introduced in order to make $E^m_\mu$ dimensionless. On the level of the composite bosonic fields $E^m_\mu$ and $g_{\mu\nu}$ the inverse vielbein
and metric $E^\mu_m(x)~,~g^{\mu\nu}(x)$ are well defined provided $E=\det(E^m_\mu)\neq 0$.
Our approach realizes the general idea that both geometry
and topology can be associated to the
properties of appropriate correlation functions \cite{GenG}.

Due to the lack of local Lorentz symmetry the global vielbein contains additional degrees of
freedom that are not described by the metric. Correspondingly,
the effective theory of gravity will also
exhibit new invariants not present in Einstein gravity. These invariants
are consistent with global but not local Lorentz symmetry.
Indeed, we may use a nonlinear field decomposition $E^m_\mu(x)=e^m_\mu(x)H_m^{\ n}(x)$ where
$e^m_\mu$ describes the usual ``local vielbein'' and $H_m^{\ n}$ the additional degrees of
freedom. These additional degrees of freedom are massless Goldstone-boson-like
excitations due to the spontaneous breaking of a global symmetry. In Einstein gravity,
$H_m^{\ n}$ would be the gauge degrees of freedom of the local Lorentz transformations and
therefore drop out of any invariant action. In contrast, the generalized gravity discussed
here will lead to new propagating massless gravitational degrees of freedom. Indeed, the
kinetic terms for $H_m^{\ n}$ can be inferred from the most general effective action
containing two derivatives which is invariant under diffeomorphisms and global Lorentz
transformations
\begin{eqnarray}\label{AA2}
\Gamma_{(2)}&=&\frac{\mu}{2}\int d^dxE\Big\{-R+\tau[D^\mu
E^\nu_mD_\mu E^m_\nu\nonumber\\
&&-2D^\mu E^\nu_m D_\nu E^m_\mu]
+\beta D_\mu E^\mu_m D^\nu E^m_\nu\Big\}.
\end{eqnarray}
Here the curvature scalar $R$ is constructed from the metric $g_{\mu\nu}$ which is also used
to lower and raise world indices in the usual way. The covariant derivative $D_\mu$
contains the connection $\Gamma_{\mu\nu}^{\ \ \lambda}$ constructed from $g_{\mu\nu}$ but
no spin connection. Due to the missing spin connection the last two terms
$\sim\tau,\beta$ are invariant under global but not local Lorentz transformations. They
induce the kinetic term for $H_m^{\ n}$. The effective action (\ref{AA2}), together with a
``cosmological constant'' term $\sim\int d^dxE$, constitutes the first order in a systematic
derivative expansion. In the one loop approximation to spinor gravity
one finds $\beta=0$.

An investigation of the gravitational field equations derived from (\ref{AA2}) in a four dimensional setting reveals \cite{SG1} that the invariant $\sim\tau$ is not constrained by anyone of the present observations of gravity. In particular, it does not modify linear gravity  or the first nontrivial order of post-Newtonian-gravity. The isotropic Schwarzschild solution and the homogeneous and isotropic cosmological solutions remain unaffected. One concludes that the {\em local} character of the Lorentz-symmetry is actually not tested by observation \cite{SG1} since an unconstrained term not invariant with respect to the {\em local} Lorentz transformations remains allowed. 

In contrast, severe bounds exist for the coupling $\beta$. We will see, however, that the term $\sim \beta$ is not present in the effective gravitational field equations that obtain as the conditions for solutions of the Schwinger-Dyson equations for spinor gravity.

In the present paper we generalize the action (\ref{AA0}) and discuss in sect. \ref{spinoraction} the construction of invariants with respect to diffeomorphisms and global Lorentz symmetry from fundamental spinors. The issue of local Lorentz symmetry is addressed in the appendix A. 

In the following sections we turn to the geometrical concepts implied by the absence
of local Lorentz symmetry. This is treated in the context of the most general form of the effective action for the gravitational degrees of freedom (sect. \ref{globallorentzsymmetry}).   We present in sect. \ref{nonlinearfields} a nonlinear decomposition of the
global vielbein $E^m_\mu$ into the usual local vielbein $e^m_\mu$ and a nonlinear field
$H_m{^n}$ which contains the new degrees of freedom. We also discuss different possible
definitions of covariant derivatives which correspond to alternative but equivalent
geometrical viewpoints. The implications of generalized gravity \cite{CWGG} for the general structure of
the solutions of the field equations are investigated in sect. \ref{generalizedgravity}.

In sect. \ref{gravitationalspinor} we address
the gravitational interactions of the fermions in our generalized setting. Global Lorentz
symmetry is sufficient in order to forbid mass terms for irreducible spinors in
$d=2,6,8,9~mod~8$ \cite{CWMS}. The absence of local Lorentz symmetry allows, however, an unconventional
gravitational coupling of the spin of the fermions. In sect. \ref{nonlinearspinorfields} we discuss the
gravitational couplings of the fermions in a nonlinear language. One sees how the absence
of the usual coupling to the totally antisymmetric part of the spin connection is equivalent
to a nonzero coupling to an antisymmetric tensor field $c_{\mu\nu}$.  This coupling would
only be observable for objects with a macroscopic spin vector. Using the nonlinear
formulation we briefly address the one loop approximation to spinor
gravity in sect. \ref{oneloopspinorgravity} and show that $\beta$ vanishes in one loop order. These findings are extended to a non-perturbative framework based on the solution of Schwinger-Dyson equations in sect. \ref{schwingerdysonequation}. Sect. \ref{conclusions} presents our conclusions.

\section{Spinor action}
\label{spinoraction}
The spinor fields $\psi(x)$ are represented by anticommuting
Grassmann variables. They transform as irreducible spinor representations under the
$d$-dimensional Lorentz group $SO (1,d-1)$ or the Euclidean orthogonal group $SO(d)$
\begin{equation}\label{1}
\delta_{{\cal L}}\psi~=~-\frac{1}{2}\epsilon_{mn}
\Sigma^{mn}\psi~,~\Sigma^{mn}=-\frac{1}{4}[\gamma^m,\gamma^n].
\end{equation}
Here the Dirac matrices obey
\begin{equation}\label{4AA}
\{\gamma^m,\gamma^n\}=2\eta^{mn},
\end{equation}
and Lorentz indices
are raised and lowered by $\eta^{mn}$ or $\eta_{mn}$. For $SO(1,d-1)$ one has
$\eta^{mn}=\eta_{mn}$ =$~diag$$~(-1,+1,\dots,+1)$, whereas for the Euclidean case
$\eta^{mn}=\eta_{mn}=\delta^n_m(\equiv \delta_{mn}\equiv \delta^{mn})$.
More generally, we will consider a signature of $\eta_{mn}$ with $s$ eigenvalues $-1$
and $d-s$ eigenvalues $+1$, with $s=0,1$ for Euclidean and Minkowski signature,
respectively. Under $d$-dimensional
general coordinate transformations the spinor fields transform as scalars
\begin{equation}\label{2}
\delta_\xi\psi=-\xi^\nu\partial_\nu\psi
\end{equation}
such that $\partial_\mu\psi$ is a vector. Similarly, the spinor fields $\bar{\psi}(x)$
transforms as
\begin{equation}\label{3}
\delta_{{\cal L}}\bar{\psi}=\frac{1}{2}\bar{\psi}\epsilon_{mn}\Sigma^{mn}~,~
\delta_\xi\bar{\psi}=-\xi^\nu\partial_\nu\bar{\psi}.
\end{equation}
For Majorana spinors one has $\bar{\psi}=\psi^TC$ where $C$
obeys \footnote{For details see \cite{CWMS}} $(\Sigma^T)^{mn}=-C\Sigma^{mn}C^{-1}$.
Otherwise $\bar{\psi}$ may be considered as an independent spinor, where an involutive
mapping between $\psi$ and $\bar{\psi}$ can be
associated to complex conjugation in spinor space.
In even dimensions the irreducible spinors are Weyl spinors obeying
$\bar{\gamma}\psi=\psi$ with $\bar{\gamma}=\eta\gamma^0\dots\gamma^{d-1},
\eta^2=(-1)^{d/2-s},\bar{\gamma}^2=1,\bar{\gamma}^\dagger=\bar{\gamma}$.

We want to construct an action that is a polynomial in $\psi,\bar{\psi}$ and invariant
under global Lorentz-transformations and general coordinate transformations. Our first
basic building block is a spinor bilinear
\begin{equation}\label{3AA}
\tilde{E}^m_\mu=i\bar{\psi}\gamma^m\partial_\mu\psi.
\end{equation}
It transforms as a vector under general coordinate
transformations
\begin{equation}\label{4}
\delta_\xi\tilde E^m_\mu=-
\partial_\mu\xi^\nu\tilde E^m_\nu-\xi^\nu\partial_\nu\tilde E^m_\mu,
\end{equation}
and as a vector under global Lorentz rotations
\begin{equation}\label{4a}
\delta_{\cal L}\tilde E^m_\mu=-\tilde E^n_\mu\epsilon_n\ ^m.
\end{equation}
The transformation properties of $\tilde{E}^m_\mu$ under {\em local} Lorentz transformations are discussed in appendix A. In general, the expectation value of $\tilde{E}^m_\mu$ may be complex.

From $\tilde E^m_\mu$ we can easily construct a composite field transforming like the metric
\begin{equation}\label{5}
\tilde g_{\mu\nu}=\tilde E^m_\mu\tilde E^n_\nu\eta_{mn}.
\end{equation}
However, no object transforming as the inverse metric can be constructed as a polynomial
in the spinor fields. The spinor polynomials contain only ``lower world indices'' $\mu,\nu$
which are induced by derivatives. The only possible coordinate invariant polynomial must
therefore involve precisely $d$ derivatives, contracted with the totally antisymmetric
$\epsilon$-tensor. In particular, the scalar density $\tilde E=\det(\tilde E^m_\mu)$ can
be written as a spinor polynomial
\begin{equation}\label{6}
\tilde E=\frac{1}{d!}\epsilon^{\mu_1\dots\mu_d}
\epsilon_{m_1\dots m_d}
\tilde E^{m_1}_{\mu_1}\dots
\tilde E^{m_d}_{\mu_d}=\det(\tilde{E}^m_\mu).
\end{equation}
Therefore a possible invariant action reads
\begin{equation}\label{7}
S_E=\alpha\int d^dx\tilde E.
\end{equation}
It involves $d$ derivatives and $2d$ powers of $\psi$.

For the discussion of
further polynomial invariants we note that the ways to construct
invariants are restricted by the absence of objects transforming as the inverse metric
or the inverse vielbein. All invariants contain $\epsilon^{\mu_1\dots\mu_d}$ where the
indices $\mu_1\dots\mu_d$ have to be contracted with derivatives. We emphasize that
the action $S_E$ is only invariant under {\it global} Lorentz rotations, but not {\it local}
Lorentz rotations. For a first classification of spinor polynomials which are invariant with respect to
diffeomorphisms and global Lorentz rotations we use as an example a
Majorana-Weyl (MW) spinor in ten dimensions which exists for a Minkowski signature. The generalization to other dimensions and other 
signature is
straightforward. Since the MW-spinors are the most restricted type of irreducible spinors
one typically finds additional invariant structures if the Majorana or Weyl constraint
is absent. We will find a large number of possible invariants. Nevertheless, this number
remains finite, as a consequence of the Grassmann nature of the spinor fields.

We begin with the construction of invariants which involve $2d$ powers of
$MW$-spinors and $d$
derivatives. Invariance under general coordinate transformations requires that with respect
to the world indices $\mu$ (associated to derivatives $\partial_\mu$) we have to construct a
totally antisymmetric tensor of rank $d$. This involves $d$ derivatives. We can always
group one spinor with a derivative and one without a derivative into a bilinear
$\psi_\alpha\partial_\mu\psi_\beta$. Such a bilinear transforms as a vector and will
be our basic building block. For $d=10$ these basic building blocks can be written in
terms of $\gamma$-matrices as
\begin{eqnarray}\label{A1}
\tilde{E}^m_\mu&=&i\bar{\psi}\gamma^m\partial_\mu\psi~~,~~\tilde{F}_\mu^{\ m_1m_2m_3}=
i\bar{\psi}\gamma^{m_1m_2m_3}_{(3)}\partial_\mu\psi,\nonumber\\
&&\tilde{G}_\mu^{\ m_1\dots m_5}=i\bar{\psi}\gamma^{m_1\dots m_5}_{(5)}
\partial_\mu\psi
\end{eqnarray}
where $\gamma_{(k)}$ are totally asymmetrized products of $k~\gamma$-matrices. Under
{\em global} Lorentz transformations these basic building blocks transform as the
appropriate totally antisymmetric tensors. All possible invariants contain 10 such
building blocks with world indices contracted by $\epsilon^{\mu_1\dots\mu_{10}}$.
The Lorentz indices can be contracted by $\epsilon_{m_1\dots m_{10}}$ and
$\eta_{mn}$.

The simplest invariant takes ten times the first building block, contracted with
$\epsilon$
\begin{equation}\label{A2}
S_1=\int d^{10}x\epsilon^{\mu_1\dots\mu_{10}}\epsilon_{m_1\dots m_{10}}
\tilde{E}^{m_1}_{\mu_1}\dots\tilde{E}^{m_{10}}_{\mu_{10}}.
\end{equation}
This corresponds to the action $S_E$ (\ref{7}) on which we will often
concentrate in this paper.
For the construction of further invariants we
may replace one or several $\tilde{E}$ by $\tilde{F}$ or $\tilde{G}$.
On the level of two factors $\tilde{E}$ four allowed index contractions can replace
\begin{eqnarray}\label{A3}
\tilde{E}^{m_1}_{\mu_1}\tilde{E}^{m_2}_{\mu_2}\rightarrow\left\{
\begin{array}{l}
\tilde{E}_{\mu_1p}\tilde{F}_{\mu_2}^{\ m_1m_2p}-(\mu_1\leftrightarrow\mu_2)\\
\tilde{F}_{\mu_1}^{\ m_1pq}\tilde{F}_{\mu_2}^{\ m_2} \ _{pq}\\
\tilde{F}_{\mu_1}^{\ pqs}\tilde{G}_{\mu_2}^{\ m_1m_2} \ _{pqs}-(\mu_1\leftrightarrow\mu_2)\\
\tilde{G}_{\mu_1}^{\ m_1pqst}\tilde{G}_{\mu_2}^{\ m_2} \ _{pqst}.
\end{array}\right.
\end{eqnarray}
On the level of two derivatives we have therefore already five basic building blocks
which all transform as second rank antisymmetric tensors with respect to general coordinate
and Lorentz transformations. Already at this level we can construct a large number of
invariants $(5^5)$ by combining five such blocks and contracting with
$\epsilon_{\mu_1\dots\mu_{10}}$ and $\epsilon^{m_1\dots m_{10}}$. Additional
possibilities arise by further possible replacements on the level of three or four
derivatives, like \\
\begin{eqnarray}\label{A.4}
&&\tilde{E}^{m_1}_{\mu_1}\tilde{E}^{m_2}_{\mu_2}\tilde{E}^{m_3}_{\mu_3}\rightarrow
\tilde{F}_{\mu_2}^{\  m_1pq}
\tilde{F}_{\mu_1}^{\  m_2}\ _{qs}
\tilde{F}_{\mu_3}^{\  m_3s} \ _p,\nonumber\\
&&\tilde{E}^{m_1}_{\mu_1}\tilde{E}^{m_2}_{\mu_2}
\tilde{E}^{m_3}_{\mu_3}\tilde{E}^{m_4}_{\mu_4}\rightarrow
\tilde{E}^p_{\mu_1}\tilde{E}^q_{\mu_2}\tilde{F}_{\mu_3}^{\ s} \ _{pq}
\tilde{G}_{\mu_4s}^{\ \ m_1m_2m_3m_4}.
\end{eqnarray}
Of course, not all of these combinations will be linearly independent. Nevertheless,
already at this level we have found a large but finite number of independent
invariants.

We next increase the number of spinor fields that may appear in the polynomials.
Since the number of derivatives is fixed to be exactly $d$ the new invariants
can be constructed by use of the non-derivative bilinears
\begin{equation} \label{A4}
\tilde{Q}^{mnp}=\frac{i}{4}\bar{\psi}\gamma^{mnp}_{(3)}\psi.
\end{equation}
In ten dimensions the 120-component antisymmetric tensor contains
all nonvanishing bilinears.
We note $\partial_\mu \tilde Q^{mnp}=\tilde{F}_\mu^{\ \ mnp}/2$.
The total antisymmetry of the $\epsilon$-tensor implies
\begin{eqnarray}\label{A9}
\epsilon_{m_1m_2\dots m_d}\tilde Q^{m_1m_2m_3}
\tilde Q^{m_4m_5m_6}&=&0~,\nonumber\\
\epsilon_{m_1\dots m_d}\tilde Q^{m_1pq}
\tilde Q^{m_2}_{\ \ \ pq}&=&0
\end{eqnarray}
whereas the combination $\epsilon_{m_1\dots m_d}
\tilde Q^{m_1m_2p}\tilde Q^{m_3m_4}_{\ \ \ \ \ \ p}$
does not vanish. The combination $\tilde{Q}^{mnp}\tilde{Q}_{mnp}$ must correspond to a
Lorentz-singlet in the totally antisymmetric product of four MW-spinors. Similarly,
$(\tilde{Q}^{mnp}\tilde{Q}_{mnp})^2)$ should be a singlet in the totally antisymmetric
product of eight spinors and so on.
Obviously, one cannot construct arbitrarily high powers of $\tilde{Q}$ since the
antisymmetric product of $n$ spinors vanishes once $n$ exceeds the dimension of the
spinor representation. The maximal $n$ corresponds to a Lorentz-singlet and is unique. It is given by the totally antisymmetric product of $D_s$ spinors, with $D_s$ the dimension of the spinor representation. For the Majorana-Weyl spinor in ten dimensions one has $D_s=16$. We conclude
that the number of independent invariants increases substantially if we allow for
polynomials with an arbitrary number of spinor fields. It remains finite, however.

We can also think about invariants with less than $2d$ spinor fields. The minimal number
for even dimensions is $d+2$. One needs
$d$ derivatives acting on $d$ different spinors. However, an invariant with only $d$
factors $\partial_\mu\psi$ is a total derivative - partial integration always leads to
two derivatives acting on a single spinor which vanish due to the antisymmetry of the
$\epsilon$-tensor. Since the action should be bosonic we have to add at least two more
spinors.

In order to study this issue we investigate as building blocks the bilinears with two
derivatives
\footnote{Due to the contraction with the $\epsilon$-tensor we can restrict the discussion
to tensors that are antisymmetric in $\mu_1\leftrightarrow \mu_2$.}
\begin{eqnarray}\label{B1}
\tilde{H}_{\mu_1\mu_2,\alpha\beta}&=&\partial_{\mu_1}\psi_\alpha\partial_{\mu_2}\psi_\beta
-\partial_{\mu_2}\psi_\alpha\partial_{\mu_1}\psi_\beta\nonumber\\
&=&\partial_{\mu_1}\psi_\beta\partial_{\mu_2}\psi_\alpha-
\partial_{\mu_2}\psi_\beta\partial_{\mu_1}\psi_\alpha.
\end{eqnarray}
The antisymmetry $\tilde{H}_{\mu_1\mu_2,\alpha\beta}=-
\tilde{H}_{\mu_2\mu_1,\alpha\beta}$ and the Grassmann anticommutation property of the
spinors requires symmetry in the spinor indices $\alpha,\beta,$ i.e. $\tilde{H}_{\mu_1\mu_2,
\alpha\beta}=\tilde{H}_{\mu_1\mu_2,\beta\alpha}$. For a Majorana Weyl spinor in $d=10$
the $16\cdot 17/2=136$ symmetric bilinears can be decomposed into two irreducible
representations of the Lorentz group
\begin{equation}\label{B9A}
\tilde{S}_{\mu_1\mu_2}^{\ \ \ m}=i(\partial_{\mu_1}\bar{\psi})\gamma^m
\partial_{\mu_2}\psi-i(\partial_{\mu_2}\bar{\psi})\gamma^m\partial_{\mu_1}\psi
\end{equation}
and
\begin{eqnarray}\label{B2}
\tilde{T}_{\mu_1\mu_2}^{\ \ \ m_1m_2m_3m_4m_5}&=&i(\partial_{\mu_1}\bar{\psi})
\gamma^{m_1m_2m_3m_4m_5}_+\partial_{\mu_2}\psi\nonumber\\
&&-(\mu_1\leftrightarrow\mu_2)
\end{eqnarray}
with
\begin{eqnarray}\label{B6}
\gamma^{m_1m_2m_3m_4m_5}_{(5)}
&=&\frac{1}{252}(\gamma^{m_1}\gamma^{m_2}\gamma^{m_3}\gamma^{m_4}\gamma^{m_5}\nonumber\\
&&-\gamma^{m_1}\gamma^{m_2}\gamma^{m_3}\gamma^{m_5}\gamma^{m_4}+\dots)
\end{eqnarray}
and
\begin{eqnarray}\label{B7}
&&\gamma^{m_1m_2m_3m_4m_5}_+=\gamma^{m_1m_2m_3m_4m_5}_{(5)}
\frac{1+\bar{\gamma}}{2}\\
&&=\frac{1}{2}\{\gamma^{m_1m_2m_3m_4m_5}_{(5)}\nonumber\\
&&+\epsilon^{m_1m_2m_3m_4m_5n_1n_2n_3n_4n_5}
\gamma_{(5)n_1n_2n_3n_4n_5}\}.\nonumber
\end{eqnarray}
The bilinears $\tilde{S}$ and $\tilde{T}$ belong to the representations $10$ and
$126$ of $SO(10)$, respectively. We note that the Weyl constraint $\bar{\gamma}\psi=\psi$
implies the relation
\begin{eqnarray}\label{B8}
&&\tilde{T}_{\mu_1\mu_2}^{\ \ \ \ m_1m_2m_3m_4m_5}\\
&=&\epsilon^{m_1m_2m_3m_4m_5n_1n_2n_3n_4n_5}
\tilde{T}_{\mu_1\mu_2n_1n_2n_3n_4n_5}\nonumber
\end{eqnarray}
such that the number of independent components (for given $\mu_1,\mu_2)$ is $252/2$.
The bilinears $\tilde{S}$ and $\tilde{T}$ transform as antisymmetric tensors with
respect to the world indices $\mu_1,\mu_2$. These building blocks
with two derivatives and two spinors can also be written as
\begin{eqnarray}\label{B9}
\tilde{S}_{\mu_1\mu_2}^{\ \ \ \ m}&=&i\partial_{\mu_1}(\bar{\psi}
\gamma^m\partial_{\mu_2}\psi)
-i\partial_{\mu_2}(\bar{\psi}\gamma^m\partial_{\mu_1}\psi)\nonumber\\
&=&\partial_{\mu_1}\tilde{E}^m_{\mu_2}-\partial_{\mu_2}\tilde{E}^m_{\mu_1}
=-2\tilde{\Omega}_{\mu_1\mu_2}^{\ \ \ \ m}
\end{eqnarray}
and
\begin{eqnarray}\label{B10}
\tilde{T}_{\mu_1\mu_2}^{\ \ \ \ m_1m_2m_3m_4m_5}&=&
\partial_{\mu_1}\tilde{G}_{\mu_2}^{\ \ m_1m_2m_3m_4m_5} \nonumber\\
&&-\partial_{\mu_2}\tilde{G}_{\mu_1}^{\ \ m_1m_2m_3m_4m_5}.
\end{eqnarray}
Let us first form a polynomial of $10$ spinors using
$5$ factors $\tilde{S}$ or $\tilde{T}$.
In this case the number of Lorentz indices is odd. We can therefore form a
third rank antisymmetric tensor which can finally be contracted with
$\tilde{Q}^{mnp}$. This construction yields the invariant with the minimal number of $d+2=12$ spinors.

A possible invariant with the maximal number of $d+D_s=26$ spinors would combine the totally symmetric product of $d$ factors $\partial_\mu\psi$ with the totally antisymmetric product of $D_s$ factors $\psi$. Its construction requires that a Lorentz singlet is contained in the totally symmetric product of $d$ spinor representations. As we have seen, this is not possible for $d=10$ since five factors of $\tilde{S}$ or $\tilde{T}$ always result in a Lorentz-tensor with an odd number of indices and therefore contain no singlet. 
We conclude that for $d=10$ the invariant spinor
polynomials must contain at least $12$ and less than $26$ spinor fields.

Global Lorentz symmetry and diffeomorphism invariance allow for many
independent polynomial invariants in the ``classical action''.
Their number is finite and one may therefore speculate
that spinor gravity is a renormalizable theory. Nevertheless, the large number of
possible independent invariants is not very attractive for a fundamental theory of all
interactions - it could hinder severely the predictivity. The number of allowed invariants depends crucially on the symmetry. For example, a
{\em local} Lorentz symmetry substantially reduces the
number of invariants \cite{CWST}. At the present stage the present form of the microscopic (or classical) spinor action remains subject to many uncertainties. We therefore concentrate in this note on those aspects that follow from general considerations independently of the details of the spinor action.

\section{Effective action with Global Lorentz Symmetry}
\label{globallorentzsymmetry}
In order to derive the effective geometrical field equations for spinor gravity one should
compute the effective action $\Gamma$ for the expectation values of the spinor bilinears, in particular for the vielbein
\begin{equation}\label{28YA}
E^m_\mu(x)=\langle\tilde{E}^m_\mu(x)\rangle.
\end{equation}
This incorporates the effect of the quantum fluctuations and the exact quantum field
equations obtain by variation of $\Gamma$. In \cite{HW1,SG1} we have presented a first
calculation based on partial bosonization and the loop expansion. In the present paper
we will supplement a non-perturbative approach based on the Schwinger-Dyson equations in sect. \ref{schwingerdysonequation}. Before proceeding to this computation we will
exploit in the following sections the general structures of their outcome which are based
purely on symmetry. Indeed, if the functional measure preserves diffeomorphism and global
Lorentz symmetry the effective action has to be invariant under the corresponding
transformation as well. We will see that the lack of local symmetry generates new
geometrical structures. The insights learned from symmetry considerations are independent of the approximation and do not require a precise specification of the classical action discussed in sect. \ref{spinoraction}.

Our generalized version of gravity
contains the metric as expected from a theory with diffeomorphism invariance. It obtains in
the usual way as a product of two vielbeins
\begin{equation}\label{S1}
g_{\mu\nu}=E^m_\mu E^n_\nu\eta_{mn}.
\end{equation}
We will concentrate here on a Minkowski signature $\eta_{mn}=diag(-1,+1,+1,\dots +1)$ and restrict the discussion to real vielbeins. From this point of view we therefore recover the standard situation for
gravity in the vielbein formulation. However, in contrast to standard gravity we find that
$E^m_\mu$ describes additional degrees of freedom. Due to the lack of local
Lorentz symmetry the vielbein will contain ``physical'' degrees of freedom not
described by the metric. This can easily be seen by investigating the most general form of
the effective action consistent with general coordinate and global Lorentz transformations.
We proceed by a derivative expansion.
In lowest order the unique invariant is
$(E=\det E^m_\mu,g=|\det g_{\mu\nu}|=E^2)$
\begin{equation}\label{S2}
\Gamma_0=\epsilon\int d^dxE=\pm\epsilon\int d^dx\sqrt{g}.
\end{equation}

In contrast to the classical action, which is formulated in terms of the spinor
Grassmann variables and therefore must be a polynomial in $\psi$, the effective action
depends on bosonic fields $E^m_\mu$.
There is no reason why the effective action should be a polynomial in
$E^m_\mu$. Whenever $E\neq 0$ we can construct the inverse vielbein
\begin{eqnarray}\label{S5}
E^{\mu_1}_{m_1}&=&\frac{1}{(d-1)!E}\epsilon^{\mu_1\dots\mu_d}\epsilon_{m_1\dots m_d}
E^{m_2}_{\mu_2}\dots E^{m_d}_{\mu_d}\nonumber\\
&=&\frac{1}{E}\frac{\partial E}{\partial E^{m_1}_{\mu_1}}
\end{eqnarray}
which obeys $E^m_\mu E^\nu_m=\delta^\nu_\mu~,~E^\mu_m E^n_\mu=\delta^n_m$.
This allows us to define the inverse metric
\begin{equation}\label{S6a}
g^{\mu\nu}=E^\mu_mE^{m\nu}~,~g^{\mu\nu}g_{\nu\rho}=\delta^\mu_\rho
\end{equation}
which can be used to raise world indices.

The most general invariant effective action involving two derivatives of the
vielbein (we assume parity invariance) reads \cite{SG1}

\begin{equation}\label{27NNA}
\Gamma_2=\mu(I_1+\tau I_2+\beta I_3)
\end{equation}
with
\begin{eqnarray}\label{27NNB}
I_1&=&-\frac{1}{2}\int d^dxE~R\big[g_{\mu\nu}[E^m_\rho]\big]\\
I_2&=&\frac{1}{2}\int d^dxE\{D^\mu E^\nu_mD_\mu E^m_\nu
-2D^\mu E^\nu_mD_\nu E^m_\mu\}\label{27NNC}\\
I_3&=&\frac{1}{2}\int d^dxED_\mu E^\mu_mD^\nu E^m_\nu.\label{27NND}
\end{eqnarray}
Here we have introduced the covariant derivative
\begin{eqnarray}\label{S10}
D_\mu E^m_\nu&=&\partial_\mu E^m_\nu-\Gamma_{\mu\nu}^{\ \ \lambda} E^m_\lambda~,\nonumber\\
D_\mu E^\nu_m&=&\partial_\mu E^\nu_m+\Gamma_{\mu\lambda}^{\ \ \nu} E^\lambda_m.
\end{eqnarray}
We emphasize that the covariant derivative acting on $E^m_\mu$ does not contain a spin
connection since $m$ is only a global Lorentz index.
As usual, the curvature scalar $R$ can be constructed from the metric $g_{\mu\nu}$
and the connection $\Gamma$ such that $R[g_{\mu\nu}]$ is a scalar, with
affine connection
\begin{equation}\label{S11}
\Gamma_{\mu\nu}^{\ \ \lambda}=
\frac{1}{2}g^{\lambda\rho}(\partial_\mu g_{\nu\rho}+\partial_\nu
g_{\mu\rho}-\partial_\rho g_{\mu\nu}).
\end{equation}

A realistic unified theory based on spinor gravity will be formulated in more than four
dimensions. One will therefore be interested in solutions of the field equations where the
characteristic length scale of $d-4$ ``internal dimensions'' is small as compared to the one
of the four dimensional ``macroscopic world''. Averaging over the not directly observable
internal space yields an effective four dimensional theory by ``dimensional reduction''.
Beyond the gravitational sector this effective action will also contain gauge interactions -
the gauge symmetries correspond to the isometries of internal space.
Nevertheless, the gravitational part of the resulting effective action will again be governed
by the symmetries of four dimensional general coordinate and global Lorentz transformations.
The systematic derivative expansion should become applicable for the large length scales
encountered in the observation of gravitational effects. We assume here that
the effective four dimensional cosmological constant is very small such that
it can be neglected for all purposes except
possibly for late cosmology. This is, of course, a highly nontrivial assumption, meaning that
spinor gravity admits an (almost) static solution with large three-dimensional characteristic
length scale (at least the size of the horizon). If so, the gravitational interactions are governed
by the effective action (\ref{27NNA}), now applied to four dimensions. 

For the four dimensional effective action it has been shown \cite{HW1}, \cite{SG1} that $\mu$ determines the effective Planck mass.
Indeed, $\mu$ and $\epsilon$ have dimension mass$^{d-2}$ and
mass$^d$, respectively. The remaining two dimensionless parameters $\tau$ and $\beta$
account for possible deviations from Einstein's gravity. We have found
\cite{SG1} that tight
observational bounds exist only for the parameter $\beta$. It is therefore
very interesting that the one loop contribution to $\beta$ vanishes.

\section{Non-linear fields}
\label{nonlinearfields}
In this paper we express the gravitational fields and the effective
action in a language somewhat closer to the usual formulation of gravity. This will shed more light on the differences between spinor gravity and Einstein gravity and on the role of global
versus local Lorentz symmetry. The role of the symmetry transformations becomes very apparent
by the use of nonlinear fields. For this purpose
we use a matrix notation where $E^m_\mu$ corresponds to the $(\mu m)$-element
of a matrix $\bar{E}$, with $E^\mu_m$ corresponding to $\bar{E}^{-1}$ and metric $g=\bar{E}\eta\bar{E}^T$. Let us introduce a pseudo-orthogonal matrix $H$ and write
$E^n_\mu=e^m_\mu H_m^{\ n}$ or
\begin{equation}\label{S12}
\bar{E}=\bar{e}H~,~H\eta H^T=\eta~,~\det H=1.
\end{equation}
The metric is independent of $H$
\begin{equation}\label{S13}
g=\bar{e}\eta \bar{e}^T~,~E=\det\bar{E}=\det \bar{e}=e.
\end{equation}
In the usual formulation of gravity with vielbein and local Lorentz invariance
the fields in $H$ are the gauge degrees of freedom of the Lorentz group. In case of local
Lorentz symmetry the effective action would be independent of $H$. In our approach
with only global Lorentz symmetry, however, we expect that $\Gamma$ depends on $H$.

The redundancy of the composition $\bar{E}=\bar{e}H$ is related to a new version
of a {\em local} Lorentz group which acts on $e$ from the right and on $H$ from the left
\begin{equation}\label{S14}
\tilde{\delta}_{\cal L}e^m_\mu=-e^n_\mu\tilde{\epsilon}_n\ ^m(x)~,~
\tilde{\delta}_{\cal L}H_m^{\ n}=\tilde{\epsilon}_m^{\ p}(x)H_p^{\ n}.
\end{equation}
Therefore $e^m_\mu$ transforms as a vector with respect to both local coordinate
and local Lorentz-transformations, just as in the usual setting. We will call $e^m_\mu$ the
``local vielbein'' and $E^m_\mu$ the ``global vielbein''. The only difference between
generalized gravity and the
standard formulation concerns the nonlinear field $H$ which transforms as a scalar with
respect to coordinate transformations and a vector with respect to the original global
Lorentz transformations
\begin{equation}\label{S15}
\delta H_m^{\ n}=-\xi^\nu(x)\partial_\nu H_m^{\ n}+\tilde{\epsilon}_m^{\ p}
(x)H_p^{\ n}-H_m^{\ p}\epsilon_p^{\ n}.
\end{equation}
Again, in case of Einstein gravity the transformations $\epsilon_p{^n}$ would be local
and permit to achieve $H_p{^n}=\delta^n_p$ without using the ``reparameterizations''
$\tilde{\epsilon}_m{^p}$. This is not possible for global parameters $\epsilon_p{^n}$.

There are two alternatives for defining the covariant derivative of the local vielbein
$e^m_\mu$. They
are connected to two different but equivalent geometrical viewpoints. The first
alternative uses the standard covariant derivative involving the spin connection $\omega[e]$
with covariantly constant local vielbein $D_\mu e^n_\nu=0$ such that
\begin{eqnarray}\label{141PA}
D_\mu E^m_\nu&=&e^n_\nu D_\mu H_n^{\ m}~,\nonumber\\
D_\mu H_n^{\ m}&=&\partial_\mu H_n^{\ m}-\omega_\mu^{\ p}{_n}[e]H_p^{\ m}.
\end{eqnarray}
This is the formulation used in \cite{HW1}.
The second alternative keeps $H_n{^m}$ covariantly constant,
$\bar{D}_\mu H_n{^m}=0$. This yields a different definition of the covariant
derivative $\bar{D}_\mu$ by
\begin{equation}\label{S15a}
D_\mu E^m_\nu=\bar{D}_\mu e^n_\nu H_n^{\ m}.
\end{equation}
Therefore the modified covariant derivative $\bar{D}_\mu$ acts on $e^n_\nu$ as
\begin{equation}\label{S17}
\bar{D}_\mu e^n_\nu=\partial_\mu e_\nu^n-\Gamma_{\mu\nu}^{\ \ \lambda}
e_\lambda^n+\bar{\omega}_\mu^{\ n}\ _pe^p_\nu
\end{equation}
where the spin connection $\bar{\omega}$ is defined as
\begin{equation}\label{S18}
\tilde{\omega}_\mu^{\ n}\ _p=[(\partial_\mu H)H^{-1}]_p^{\ n}=
\partial_\mu H_p^{\ q}H^n_{\ q}=-H_p^{\ q}\partial_\mu H^n_{\ q}.
\end{equation}
(Note $H^m_{\ p}H_n^{\ p}=H_p^{\ m}H^p_{\ n}=\delta^m_n~,~H^m_{\ p}=(H^{-1})_p^{\ m}$ and the
antisymmetry $\bar{\omega}_{\mu np}=-\bar{\omega}_{\mu pn}$.) We observe that
$\bar{\omega}_{\mu np}$ is invariant under the global Lorentz-transformations
$\delta_{\cal L}$
whereas it receives the inhomogeneous piece of a connection with respect
to the local Lorentz transformations
\begin{equation}\label{S19}
\tilde{\delta}_{\cal L}\bar{\omega}_{\mu np}=\tilde{\epsilon}_n^{\ s}
\bar{\omega}_{\mu sp}
+\tilde{\epsilon}_p^{\ s}\bar{\omega}_{\mu ns}-\partial_{\mu}\tilde{\epsilon}_{np}.
\end{equation}
Therefore $\bar{D}_\mu e_\nu^n$ transforms as a tensor both with respect to
coordinate and local
Lorentz transformations. The covariant derivative $\bar{D}_\mu$ is metric, i.e.
\begin{equation}\label{146XA}
\bar{D}_\mu g_{\nu\rho}=\bar{D}_\mu e^n_\nu e_{\rho n}
+\bar{D}_\mu e^n_\rho e_{\nu n}=0.
\end{equation}

The geometrical aspects of the covariant derivative $D_\mu$ (\ref{141PA}) are described
in \cite{HW1}. Using this formulation, one finds for the invariants
\begin{eqnarray}\label{48XA}
I_2&=&\frac{1}{2}\int d^dxe\{D^pH^{nm}D_pH_{nm}
-2D^pH_{nm}D^nH_p\ ^m\}\nonumber\\
I_3&=&\frac{1}{2}\int d^dxeD^nH_{nm}D^pH_p\ ^m
\end{eqnarray}
with $D^pH^{nm}=e^{p\rho}D_\rho H^{nm}$ etc. These invariants provide the kinetic terms
for the massless field $H_{mn}$. Their absence for Einstein gravity reflects directly
the fact that $H_{mn}$ are gauge degrees of freedom in this case.

We next turn to the viewpoint of the covariant
derivative $\bar{D}_\mu$ (\ref{S17}).
The covariant derivative $\bar{D}_\mu e^n_\nu$ defines an additional tensor structure
\cite{CWGG}. This is connected to the new invariants involving two derivatives which
are allowed in spinor gravity but not in Einstein gravity. Indeed, we may express
the effective action in terms of the nonlinear fields $e$ and $H$.
In particular, one has
\begin{eqnarray}\label{S16}
S_{\mu\nu}\ ^m&=&\partial_\mu E^m_\nu-\partial_\nu E^m_\mu\nonumber\\
&=&(\partial_\mu e^n_\nu -\partial_\nu e^n_\mu)
H_n^{\ m}+e^n_\nu\partial_\mu H_n^{\ m}-e^n_\mu\partial_\nu H_n^{\ m}\nonumber\\
&=&(\bar{D}_\mu e_\nu^n-\bar{D}_\nu e^n_\mu)H^m_n
\end{eqnarray}
and the invariants $I_k$ in (\ref{27NNB}-\ref{27NND})
can be written in terms of covariant derivatives
\begin{eqnarray}\label{S20}
I_1&=&-\frac{1}{2}\int d^dx e~R\big[g[e]\big],\nonumber\\
I_2&=&\frac{1}{2}\int d^dxe\{\bar{D}^\mu e^\nu_m
\bar{D}_\mu e^m_\nu -2\bar{D}^\mu e^\nu_m\bar{D}_\nu e^m_\mu\}\nonumber\\
I_3&=&\frac{1}{2}\int d^dx e(\bar{D}_\mu e^\mu_m)(\bar{D}^\nu e_\nu^m),
\end{eqnarray}
demonstrating the role of the tensor $\bar{D}_\mu e^n_\nu$. We recall that
in this formulation the covariant derivative
of the vielbein does not vanish. Such a situation was investigated systematically in
the generalized gravity of \cite{CWGG}.

Finally, we note that for $d=4$ there exists also a
parity violating invariant
\begin{eqnarray}\label{S23}
\Gamma_{2,p}&=&\frac{1}{4}\int d^4x\epsilon^{\mu_1\dots\mu_4}S_{\mu_1\mu_2}^{\ \ \ \ m}
S_{\mu_3\mu_4}^{\ \ \ \ n}
\eta_{mn}\nonumber\\
&=&\int d^4x\epsilon^{\mu_1\dots\mu_4}\bar{D}_{\mu_1}e^m_{\mu_2}\bar{D}_{\mu_3}e_{\mu_4m}.
\end{eqnarray}
It is obvious that in higher orders in the derivative expansion the invariants can be
constructed using as building blocks the curvature tensor $R_{\mu\nu\sigma\lambda}[e]$
and the covariant derivative $\bar{D}_\mu e^m_\nu$. The dependence on $H$ always involves
derivatives of $H$ and appears in the form of $\bar{D}_\mu e^m_\nu$.

\section{Generalized gravity}
\label{generalizedgravity}
Spinor gravity admits different geometrical viewpoints, depending on the choice of covariant
derivatives and connections. They all can be described within the framework
of generalized gravity \cite{CWGG}. In \cite{CWGG} it has been shown that the dimensional
reduction of generalized gravity has important consequences for the issue of massless
chiral spinors and it is therefore useful to make direct contact with this work.
In terms of the nonlinear fields $e^m_\mu$ and $H_n^{\ p}$ the bosonic effective action
$\Gamma[e,H]$ can be constructed from the curvature tensor $R_{\mu\nu\sigma\lambda}[e]$
and from the tensor
\begin{equation} \label{G1}
U_{\mu\nu m}=\bar{D}_\mu e_{\nu m}.
\end{equation}
The covariant conservation of the metric (\ref{146XA}) implies the antisymmetry of
$U_{\mu\nu\rho}$  in the last two indices
\begin{equation}\label{151XA}
U_{\mu\nu\rho}=e^m_\rho U_{\mu\nu}^{\ \ m}=-U_{\mu\rho\nu}.
\end{equation}
In turn, the connection $\bar{\omega}_{\mu np}[H]$ (eq. (\ref{S18})) can
be expressed as the sum
of the usual spin connection $\omega_{\mu np}[e]$ and a tensor $K_{\mu np}$,
\begin{equation} \label{G2}
\bar{\omega}_{\mu np}[H]=\omega_{\mu np}[e]+K_{\mu np},
\end{equation}
where
\begin{eqnarray}\label{G3}
\omega_{\mu np}[e]&=&e^m_\mu\omega_{mnp}[e]~~,\nonumber\\
\omega_{mnp}[e]&=&-\Omega_{mnp}+\Omega_{npm}-\Omega_{pmn}~,~\nonumber\\
\Omega_{mnp}&=&-\frac{1}{2}(e^\mu_me^\nu_n-e^\mu_ne^\nu_m)\partial_\mu e_{\nu p}.
\end{eqnarray}
This follows from the fact that $\omega_{\mu np}[e]$ has the inhomogeneous
transformation property of a connection (the same as $\bar{\omega}_{\mu np}[H]$ in
eq.(\ref{S19})). Hence the difference between two connections transforms as a tensor.

The tensor $K_{\mu np}$ can be expressed in terms of the tensor $U_{\mu\nu m}$ by
\begin{eqnarray} \label{G4}
K_{\mu np}&=&\frac{1}{2}e^m_\mu\{U_{mnp}-U_{nmp}-U_{npm}\nonumber\\
&+&U_{pnm}+U_{pmn}-U_{mpn}\},\nonumber\\
U_{pnm}&=&e^\mu_pe^\nu_nU_{\mu\nu m}.
\end{eqnarray}
We note that the indices of $K$ and $U$ are changed by multiplication with the local
vielbein $e^m_\mu$. In particular, the antisymmetric parts are equal tensors
\begin{equation} \label{155MA}
K_{[\mu\nu\rho]}=U_{[\mu\nu\rho]}.
\end{equation}
Similarly, $K_{[mnp]}=U_{[mnp]}$ are scalars that transform as totally antisymmetric
tensors under the {\em local} Lorentz transformations. On the other hand, eq.
(\ref{G3}) implies
\begin{equation} \label{155MB}
\omega_{[mnp]}=-\Omega_{[mnp]}.
\end{equation}
These latter quantities are scalars under general coordinate transformations but have
the inhomogeneous transformation properties of a connection with respect to the local
Lorentz transformations. The two sets (\ref{155MA}) and (\ref{155MB}) are related by
eq. (\ref{G2}). They coincide for a suitable Lorentz gauge, i. e.
$H=1, \bar{\omega}=0$.

The possible differences between solutions of generalized gravity and Einstein gravity are
intimately connected to a nonvanishing $U_{\mu\nu m}$. Indeed, we will show that any
vielbein $e^m_\mu$ which solves Einstein's equation is also a solution of generalized
gravity provided that there exists a suitable $H_0[e]$ such that $U[e,H_0]$ vanishes. Then,
in turn, one can infer from $K_{\mu np}=0$ that for these solutions
the spin connection becomes the standard
expression $\bar{\omega}_{\mu np}=\omega_{\mu np}[e]$.

For this purpose we express the effective action (\ref{27NNA}) in terms of $U_{\mu\nu m}$
\begin{eqnarray} \label{G5}
\Gamma&=&\mu\int d^dx\det e\nonumber\\
\{&-&\frac{1}{2} R[e]
+\frac{3}{2}\tau U_{[\mu\nu\rho]}U^{[\mu\nu\rho]}
+\frac{1}{2}\beta U_\mu^{\ \mu}\ _mU_\nu^{\ \nu m}\}.
\end{eqnarray}
The field equation in absence of sources
\begin{equation} \label{G6}
\frac{\delta\Gamma}{\delta E^m_\mu}=0
\end{equation}
is obeyed for
\begin{equation} \label{G7}
\frac{\delta\Gamma[e,H]}{\delta e^m_\mu}=0~,~\frac{\delta\Gamma[e,H]}{\delta H_n^{\ p}}=0.
\end{equation}
We start with the observation that $\Gamma$ depends on $H$ only via $U$ and solve first
the second equation as a functional of $e$
\begin{equation} \label{G8}
\frac{\delta\Gamma}{\delta H_n^{\ p}}=
\frac{\delta\Gamma[e,U]}{\delta U_{\mu\nu m}\ _{|e}}
\frac{\delta U_{\mu\nu m}}{\delta H_n^{\ p}\ _{|e}}=0.
\end{equation}
Since the effective action (\ref{G5}) is quadratic in $U$ eq.(\ref{G8}) has
a solution
\begin{equation} \label{G9}
U^0_{\mu\nu m}\big[e,H_0[e]\big]=0
\end{equation}
provided that a suitable $H_0[e]$ exists. We next reinsert this solution into the
effective action and define
\begin{equation} \label{G10}
\Gamma[e]=\Gamma\big[e,H_0[e]\big]=\Gamma\Big[e,U^0\big[e,H_0[e]\big]\Big].
\end{equation}
Any solution of the remaining field equation
\begin{equation} \label{G11}
\frac{\delta\Gamma[e]}{\delta e^m_\mu}=0
\end{equation}
also solves the first equation in (\ref{G7}) and is therefore
a solution of the original field equation (\ref{G6}). We finally observe that $\Gamma[e]$
contains only the term $\sim R$ and can therefore be expressed in terms of the metric. The
field equation therefore becomes
\begin{equation} \label{G12}
\frac{\delta\Gamma[g]}{\delta g^{\mu\nu}}=0
\end{equation}
which is precisely the Einstein equation
\begin{equation} \label{G13}
R_{\mu\nu}-\frac{1}{2}R g_{\mu\nu}=0.
\end{equation}

This concludes the proof of our assertion that a vielbein which solves eqs.
(\ref{G13}) and (\ref{G9}) also solves the field equation (\ref{G6}) of generalized
gravity. More generally, the effective action of generalized gravity contains terms that
can be constructed from the curvature tensor $R_{\mu\nu\sigma\lambda}$ (and therefore only
involve $g_{\mu\nu}$) plus terms with at least two powers of $U$. For fields with
$U=0$ the contribution of the latter terms to the field equations vanishes.

Let us next investigate what
are the conditions for the existence of a suitable
configuration $H_0[e]$ such that $U^0=0$. According to eqs. (\ref{G2}), (\ref{G4}) this
requires
\begin{equation} \label{G14}
\partial_\mu H_p^{\ q}H_{nq}=\omega_{\mu np}[e].
\end{equation}
This condition is equivalent to the one that for a given vielbein $e_\mu^n$ there
exists a matrix $H_n^{\ p}$ such that $E_\mu^p=e^n_\mu H_n^{\ p}$ obeys
$D_\mu E^p_\nu=0$, as can be seen from eq. (\ref{S15a}). The question of the
existence of $H_0[e]$ therefore reduces to the question for which $e^m_\mu$ one can find
a field $H_0[e]$ such that $E^m_\mu$ has a vanishing covariant derivative
$D_\mu E^m_\nu=0$. We observe that $H_0$ acts as a local Lorentz transformation
on $e$ (eq. (\ref{S12})). We also know the identity
\begin{equation} \label{G15}
D_\mu e^m_\nu=\partial_\mu e^m_\nu-\Gamma_{\mu\nu}^{\ \ \ \lambda}
e^m_\lambda+\omega_\mu^{\ m}\ _p[e]e_\nu^p=0.
\end{equation}
(This corresponds to the vanishing of the covariant derivative if $\omega$ instead of
$\bar{\omega}$ is used and follows directly from the definition (\ref{G3}).) The identity
(\ref{G15}) is invariant under local Lorentz transformations. We therefore can achieve
$D_\mu E^m_\nu=D_\mu(e^n_\nu H_n^{\ m})=0$ provided that there exists a
Lorentz transformation $H$ by which the spin connection
$\omega[E]=\omega[eH]$
can be made to vanish. (This is precisely the content of the relation (\ref{G14}).)
We may call a connection $\omega[e]$ obeying eq.(\ref{G14}) ``Lorentz flat''. Lorentz flatness
is realized only for a special class of vielbeins. Lorentz flat solutions of Einstein
gravity remain exact solutions for generalized gravity.

General solutions for $E^m_\mu$ may not obey $U_{\mu mn}=0$. In this case we may
find deviations from standard gravity. According to eq. (\ref{G4}) they can all be written
in terms of the tensor
\begin{equation}\label{160AA}
K_{mnp}=e^\mu_mK_{\mu np}.
\end{equation}
By construction $K_{mnp}$ transforms as a scalar under general coordinate transformations.
It is a third rank tensor with respect to the local Lorentz-transformations
$\delta_{\tilde{{\cal L}}}$ (\ref{S14}), (\ref{S19}) and invariant with respect to the
global rotations $\delta_{{\cal L}}$. The tensor can be decomposed into a totally
antisymmetric part $K_{[mnp]}$ and a vector $K_m=K_{mn}^{\ \ \ n}$. We will see
(sect. \ref{oneloopspinorgravity}) that $K_m$ does not appear in the one loop approximation.

At this point we emphasize that
Lorentz-flatness $(U_{\mu np}=0~,~K_{mnp}=0)$
is a sufficient but not a necessary condition for the coincidence of solutions of
generalized gravity with Einstein gravity. This is demonstrated by the isotropic and
cosmological solutions for $\beta=0$ \cite{SG1}. Indeed, for $\beta=0$ the condition for the
coincidence of solutions of generalized gravity and Einstein gravity gets weakened:
Every solution of Einstein gravity is also a solution  of generalized gravity provided
that the totally antisymmetric part of $U^0$ vanishes
\begin{equation}\label{168PA}
U^0_{[\mu\nu\rho]}\big[e,H_0[e]\big]=0.
\end{equation}
The condition (\ref{168PA}) is substantially weaker than the condition for Lorentz flatness
(\ref{G9}). It is obeyed for a large class of solutions of Einstein gravity,
including the Schwarzschild and
Friedmann solutions \cite{HW1}. The proof proceeds in complete
analogy to the above, noting that
only $U_{[\mu\nu\rho]}$ appears in the effective action (\ref{G5}) for $\beta=0$.

We finally note that our geometrical setting can also be interpreted
in terms of torsion. Indeed, we may define a different connection
$\tilde{\Gamma}_{\mu\nu}^{\ \ \ \lambda}$ and, in consequence, a further covariant derivative $\tilde{D}_\mu$, such that the global vielbein is covariantly conserved
\begin{equation}\label{55QA}
\tilde{D}_\mu E^m_\nu=\partial_\mu E^m_\nu-\tilde{\Gamma}_{\mu\nu}^{\ \ \ \lambda}
E^m_\lambda=0.
\end{equation}
This fixes the connection as
\begin{equation}\label{55QB}
\tilde{\Gamma}_{\mu\nu}^{\ \ \ \lambda}=(\partial_\mu E^m_\nu)E^\lambda_m
\end{equation}
and comparison with eq. (\ref{S10}) identifies \footnote{
Since the l.h.s. of eq. (\ref{55QC}) is a tensor this shows that
$\tilde{\Gamma}_{\mu\nu}{^\lambda}$ indeed transforms as a connection under general
coordinate transformations. Of course, this can also be checked by direct computation
from the analogue of eq. (\ref{4}).} the contorsion
\begin{equation}\label{55QC}
E^\lambda_mD_\mu E^m_\nu=\tilde{\Gamma}_{\mu\nu}^{\ \ \ \lambda}
-\Gamma_{\mu\nu}^{\ \ \ \lambda}
=L_{\mu\nu}^{\ \ \ \lambda}.
\end{equation}
We note that the antisymmetric part of $\tilde{\Gamma}_{\mu\nu}^{\ \ \ \lambda}$ is
the torsion tensor \footnote{In ref. \cite{CWGG} the torsion tensor $T_{\mu\nu\rho}$ is
denoted by $R_{\mu\nu\rho}$}
\begin{equation}\label{55QD}
\tilde{\Gamma}_{\mu\nu}^{\ \ \ \lambda}-\tilde{\Gamma}_{\nu\mu}^{\ \ \ \lambda}
=-2\Omega_{\mu\nu}^{\ \ \ m}E^\lambda_m
=T_{\mu\nu}^{\ \ \ \lambda}.
\end{equation}
In particular, the invariant $I_3$ involves a contraction of the contorsion tensor $L$
according to the identity
\begin{equation}\label{55QE}
D^\mu E^m_\mu=(\tilde{\Gamma}_\mu^{\ \ \mu\lambda}-\Gamma_\mu^{\ \ \mu\lambda})
E^m_\lambda.
\end{equation}
Since eq. (\ref{55QA}) implies the existence of
$d$ covariantly conserved vector fields the connection
$\tilde{\Gamma}$ (\ref{55QB}) is curvature free.

Similarly, we may use in the nonlinear language a different connection and definition of
the covariant derivative such that $e^m_\mu$ is covariantly conserved even though we use the
spin connection $\bar{\omega}$. This employs a connection with torsion
\begin{equation}\label{167QA}
\tilde{D}^\prime_\mu e^n_\nu=\partial_\mu e^n_\nu-\tilde{\Gamma}_{\mu\nu}^{\ \ \lambda}
e^n_\lambda+\bar{\omega}_\mu^{\ n}~_pe^p_\nu=0
\end{equation}
and we identify
\begin{equation}\label{167QB}
U_{\mu\nu}^{\ \ \ m}=(\tilde{\Gamma}_{\mu\nu}^{\ \ \ \lambda}-\Gamma_{\mu\nu}^{\ \ \ \lambda})
e^m_\lambda=D_\mu E^m_\nu(H^{-1})_n^{\ m}
\end{equation}
in accordance with eq. (\ref{55QC}). Of course, the three formulations based on the
covariant derivatives $D_\mu,\bar{D}_\mu$ and $\tilde{D}^\prime_\mu$
(\ref{G15}) (\ref{S17}) (\ref{167QA}) are all equivalent. They can be interpreted as
different geometrical viewpoints of generalized gravity.

\section{Gravitational spinor interactions}
\label{gravitationalspinor}
In this section we turn to the gravitational interactions of the spinor fields.
They describe how fermions propagate in a gravitational background and determine
the gravitational fields created by fermionic sources. We concentrate on interactions
involving two spinor fields and use a Minkowski signature for $\eta_{mn}$.
For this purpose we use an effective action depending on both
fermionic and bosonic fields.
This amounts to the inclusion of fermionic and bosonic sources $\bar{\eta}$ and $J^\mu_m$.
The effective action is defined by an appropriate Legendre transform
\begin{equation}\label{F1}
\Gamma[\psi,E^m_\mu]=-W[\bar{\eta},J^\mu_m]+
\int d^dx\{\bar{\eta}\psi+J^\mu_mE^m_\mu\}
\end{equation}
of the logarithm of the partition function $W=\ln Z$ (cf. \cite{SG1} for details of
our setting). We need a computation of $\Gamma$ in quadratic order in $\psi$.
We concentrate on the general structure
and admit all terms consistent with the symmetries of general coordinate transformations
and global Lorentz rotations, $\Gamma_{(2)}=\int d^dx\sum_{\alpha}{\cal L}_{\alpha}$. We
start with the non-derivative interactions. For irreducible spinors in $d$ dimensions the
symmetries permit at most one possible scalar bilinear which plays the role of a fermion mass
term
\begin{equation}\label{F2}
{\cal L}_m=im_\psi E\bar{\psi}\psi.
\end{equation}
For Majorana spinors we use $\bar{\psi}=\psi^TC$. We observe that for $d=8,9 ~ mod ~ 8$
the charge conjugation matrix $C$ is symmetric, $C=C^T$. Then the
antisymmetry under the exchange of Grassmann variables forbids a mass term.
On the other side, Weyl spinors admit a mass term only for
$d=4~mod~4$. We conclude \cite{CWMS} the absence of any mass term except for
$d=3,4,5,7~mod~8$, i.e.
\begin{equation}\label{F3}
{\cal L}_m=0~\textup{for}~d=2,6,8,9~mod~8.
\end{equation}

This has a very important consequence: spinor gravity in $d=2,6,8,9~mod~8$ dimensions
necessarily leads to a spectrum containing both massless fermions and a massless
graviton! (In addition, it also involves
other massless bosonic bound states.) To be more specific, the
global Lorentz symmetry implies massless
spinors for {\em flat} $d$-
dimensional space whereas mass terms may be induced by ``compactification'' of some
of the ``internal'' dimensions or, more generally, by spontaneous breaking of some
of the symmetries in the ground state. Indeed, our statement is based purely on
symmetry: massless fermions are guaranteed by the invariance of $\Gamma$ under
global Lorentz-rotations and diffeomorphisms, whereas the absence of the mass term of the
graviton follows from the invariance under general coordinate transformations alone.
Whenever the effective action for the vielbein contains an appropriate
kinetic term with two derivatives acting on $E^m_\mu$ spinor gravity describes a propagating
massless graviton. It arises as a bound state
of massless spinors $(d=2,6,8,9~mod~8)$ or possibly massive spinors $(d=3,4,5,7~mod~8)$.
We will argue in sects. \ref{oneloopspinorgravity}, \ref{schwingerdysonequation}
that the appropriate invariant kinetic term for the
metric is indeed generated by the fluctuation effects. We also mention that our symmetry
argument only holds provided that the relevant global symmetry $(SO(1,d-1)$ or an
appropriate subgroup thereof) is free of anomalies.

As noted in \cite{CWMS} the absence of a mass term is a necessary condition for a
realistic spectrum of quarks and leptons in the effective four dimensional theory
after dimensional reduction \footnote{In presence of ``branes'' this statement holds
only if the quarks and leptons correspond to normalizable wave functions in the ``bulk''.
This seems to be a generic situation for branes with codimension larger than one where
the fermionic wave functions may be concentrated on the brane but are extended into the
bulk \cite{SW}.}. ``Spontaneous compactification'' is associated typically with a reduced symmetry
of the ground state. The chirality index \cite{CWF} determines the number of chiral fermion
generations (with quantum numbers according to the ground state symmetry) in dependence
on the properties of the internal space. More precisely, it depends on the
topology and the symmetry.
The internal space may have singularities which can be interpreted \cite{SW} as branes
\cite{CWB} or generalized black holes \cite{RS} and may be ``wharped'' \cite{RS,CWB} -
in this case the index also depends on the behavior of the geometry near the
singularity \cite{CWGG}. A small electroweak symmetry breaking at the Fermi scale will then induce a small mass
for the otherwise massless quarks and leptons \cite{CWFA}.

We next turn to the interactions involving no derivatives of the spinors and one
derivative of the vielbein. They must be of the form
\begin{equation}\label{F4}
{\cal L}_a=iE\bar{\psi}\gamma^{m_1\dots m_p}_{(p)}\psi
A^{(p)}_{m_1\dots m_p}
\end{equation}
with $A^{(p)}$ transforming as a scalar under diffeomorphisms and a totally
antisymmetric tensor under Lorentz rotations. The only tensor quantities which can
be formed from the vielbein involving only one derivative are based on
\begin{equation}\label{F5}
\Omega_{mnp}[E^m_\mu]=-\frac{1}{2}E^\mu_mE^\nu_n
(\partial_\mu E_{\nu p}-\partial_\nu E_{\mu p})
\end{equation}
which can be decomposed into irreducible representations $\Omega_m=\Omega_{mn}^{\ \ \ n}$
and $\Omega_{[mnp]}$. This yields
\begin{equation}\label{F6}
{\cal L}_a=-\frac{i}{4}E Z_a\Omega_{[mnp]}\bar{\psi}\gamma^{mnp}_{(3)}\psi
-iEZ_b\Omega_m\bar{\psi}\gamma^m\psi.
\end{equation}
We observe $\bar{\psi}\gamma^m\psi\equiv 0$ for Majorana spinors in $d=2,3,9~mod~8$
\cite{CWMS}. Furthermore, hermiticity requires $EZ_a\Omega_{[mnp]}$ to be real whereas
$EZ_b\Omega_m$ should be purely imaginary. We will
omit the term $\sim Z_b$ in the following.

Finally, we include the spinor kinetic term with one derivative acting on
the spinors $\psi$
\begin{eqnarray}\label{F7}
{\cal L}_{kin}&=&\frac{i}{2}EZ_\psi(\bar{\psi}\gamma^mE^\mu_m\partial_\mu\psi
-\partial_\mu\bar{\psi}\gamma^mE^\mu_m\psi) \nonumber\\
&&+iE\tilde{Z}_\psi E^\mu_m\partial_\mu
(\bar{\psi}\gamma^m\psi).
\end{eqnarray}
Again, the term $\sim\tilde{Z}_\psi$ is absent for Majorana spinors in
$d=2,3,9~mod~8$ and will be omitted from our discussion.
Summarizing the various terms the effective spinor action with up to one derivative reads
\begin{equation}\label{F8a}
\Gamma_{(2)}=\int d^dxE\{Z_\psi\bar{\psi}\gamma^\mu\partial_\mu\psi
-\frac{i}{4}Z_a\Omega_{[mnp]}
\bar{\psi}\gamma^{mnp}_{(3)}\psi\}.
\end{equation}
We emphasize that for irreducible spinors in $d=2,9~mod~8$ eq.
(\ref{F8a}) is the most general form consistent
with the symmetries. In the classical approximation after partial bosonization one has
$Z_\psi=-\tilde{\alpha}~,~Z_a=0$. \cite{SG1} On the other hand, the gravitational spinor
interaction in standard gravity is characterized by local Lorentz invariance which
requires $Z_a=Z_\psi$. Possible deviations from standard gravity are therefore related
to the difference $Z_\psi-Z_a$.

\section{Nonlinear spinor fields}
\label{nonlinearspinorfields}
In order to bring our formulation as close as possible to standard gravity and to make
the differences transparent we use again the nonlinear field decomposition of
sect. \ref{nonlinearfields}. We define \footnote{There should be no confusion with the
spinor bilinear $\tilde{H}$ used in sect. \ref{spinoraction}.}
\begin{equation}\label{F9a}
\psi=\tilde{H}\tilde{\psi}
\end{equation}
where $\tilde{H}$ is the appropriate spinor representation of the pseudoorthogonal matrix
$H$ in eq. (\ref{S12}). (For small deviations from unity $H_m^{\ \ n}=
\delta^n_m+\delta H_{mp}\eta^{pn}$ one has $\tilde{H}=1+\frac{1}{2} \delta H_{mn}
\Sigma^{mn}$). Using
\begin{equation}\label{F10a}
\bar{\psi}\gamma^m\psi=\bar{\tilde{\psi}}\tilde{H}^{-1}\gamma^m\tilde{H}\tilde{\psi}
=\bar{\tilde{\psi}}\gamma^n\tilde{\psi}H_n^{\ \ m}
\end{equation}
one finds
\begin{equation}\label{F11a}
\bar{\psi}\gamma^mE^\mu_m\psi=\bar{\tilde{\psi}}\gamma^m e^\mu_m\tilde{\psi}
\end{equation}
and
\begin{eqnarray}\label{F12a}
\bar{\psi}\gamma^\mu\partial_\mu\psi&=&\bar{\tilde{\psi}}\gamma^me^\mu_m
\big(\partial_\mu+\tilde{H}^{-1}(\partial_\mu\tilde{H})\big)\tilde{\psi}\nonumber\\
&=&\bar{\tilde{\psi}}\gamma^me^\mu_m(\partial_\mu-\frac{1}{2}\bar{\omega}_{\mu np}
[H]\Sigma^{np})\tilde{\psi}.
\end{eqnarray}
Here we have used the identity
\begin{equation}\label{F13a}
\tilde{H}^{-1}\partial_\mu\tilde{H}=\frac{1}{2}
\partial_\mu H_n^{\ q}H^{-1}_{qp}\Sigma^{np}
\end{equation}
and the definition (\ref{S18}) for $\bar{\omega}[H]$. Furthermore, one has
\begin{eqnarray}\label{F14a}
\Omega_{[mnp]}[E]&=&-(\omega_{[m^\prime n^\prime p^\prime]}[e]\nonumber\\
&&-\bar{\omega}_{[m^\prime n^\prime p^\prime]}[H])
H^{m^\prime}_{\ \ m}
H^{n^\prime}_{\ \ n}
H^{p^\prime}_{\ \ p}~,~\nonumber\\
\bar{\omega}_{mnp}[H]&=&e^\mu_m\bar{\omega}_{\mu np}[H].
\end{eqnarray}
In the nonlinear language the effective spinor action therefore reads
\begin{eqnarray}\label{F15}
\Gamma _{(2)}&=&\int d^dxeZ_\psi
\bar{\tilde{\psi}}\Big\{i\gamma^me^\mu_m\left(\partial_\mu
-\frac{1}{2}\bar{\omega}_{\mu np}[H]\Sigma^{np}\right)\nonumber\\
&&+\frac{i}{4}\frac{Z_a}{Z_\psi}\big(\omega_{[mnp]}[e]
-\bar{\omega}_{[mnp]}[H]\big)
\gamma^{mnp}_{(3)}\Big\}\tilde{\psi}.
\end{eqnarray}
In the field basis $(\tilde{\psi},e,H)$ we see that the only difference from the
gravitational spinor interactions in Einstein gravity involves
\footnote{For $\bar{\psi}\gamma^m\psi\neq 0$ additional couplings $\sim K_{mn}^{\ \ \ n}
\gamma^m$ are possible if $E^m_\mu$ is complex.} the tensor $K_{\mu np}$ defined
in eq. (\ref{G2})
\begin{eqnarray}\label{F16}
\Gamma _{(2)}&=&\int d^dxeZ_\psi
\bar{\tilde{\psi}}\Big\{i\gamma^me^\mu_m\left(\partial_\mu
-\frac{1}{2}\omega_{\mu np}[e]\Sigma^{np}\right)\nonumber\\
&&+\frac{i}{4}\left(1-\frac{Z_a}{Z_\psi}\right)
K_{[mnp]}\gamma^{mnp}_{(3)}\Big\}\tilde{\psi}.
\end{eqnarray}

For an interpretation of the modification of the
gravitational spinor coupling due to the term $\sim K_{[mnp]}$ it is instructive to
investigate first weak gravitational effects.
In the linear approximation for weak gravity (cf. \cite{SG1})
we can choose a parameterization where
\begin{equation}\label{F17}
e^m_\mu=\delta^m_\mu+\frac{1}{2}h_{\mu\nu}\eta^{\nu m}~,~H_m^{\ n}
=\delta^n_m+\frac{1}{2}a_{\mu\nu}\delta^\mu_m\eta^{\nu n}
\end{equation}
such that
\begin{equation}\label{F18}
\omega_{[\mu\nu\rho]}[e]=0~,~K_{[\mu\nu\rho]}=\bar{\omega}_{[\mu\nu\rho]}
[H]=-\frac{1}{2}\partial_{[\mu}c_{\nu\rho]}.
\end{equation}
By partial integration the coupling of the antisymmetric tensor field $c_{\nu\rho}$ to
spinors can also be written as
\begin{equation}\label{F19}
{\cal L}_c=\frac{1}{4}(Z_\psi-Z_a)c_{\nu\rho}\bar{\tilde{\psi}}
\gamma^{\mu\nu\rho}_{(3)}\partial_\mu\tilde{\psi}
\end{equation}
If such a coupling would be present in the effective four dimensional theory of gravity
a local fermionic particle would act as a source for $c_{\nu\rho}$, i.e.
$j_{\nu\rho}\sim\partial_\mu\bar{\tilde{\psi}}\gamma^{\mu\nu\rho}\tilde{\psi}$.
Since $c_{\nu\rho}$ is an additional massless field, Newton's law would be modified
by an additional very weak dipole-dipole interaction from the derivative couplings  $\sim(j_{\nu\rho})^2$. We note that $j_{\nu\rho}$ reflects the particular properties of
spin $1/2$ particles. Such a ``spin-contribution'' to the effective gravitational
interaction is much too weak \cite{SG1} to be seen by present experiments.

We conclude that spinor gravity leads to modified gravitational interactions of the
fermionic particles - but those modifications are not observable (at present) by
macroscopic gravity. Since the discussion of sects. \ref{gravitationalspinor},
\ref{nonlinearspinorfields} is purely based on symmetry this finding seems quite
robust as long as the symmetries (and the reality of the vielbein) admit a modification
of standard gravity only through the totally antisymmetric tensor $K_{[mnp]}$
(\ref{F8a}, \ref{F16}).

\section{Spinor gravity in one loop order}
\label{oneloopspinorgravity}
In this and the next section we turn to the derivation of the effective gravitational field equations for $E^m_\mu$ or the corresponding effective action $\Gamma$. Since our problem has no small expansion parameter there seems to be no reason why a loop expansion should be reliable. As a first non-perturbative approach for models with interacting fermions one may look for the solution of the Schwinger-Dyson equation \cite{SDE} in lowest order. This will be done in sect. \ref{schwingerdysonequation}. The structural elements of the solution to the lowest order schwinger-Dyson equation are very close to a one loop expansion. Therefore we first explore the latter in the present section. Our main finding, namely that the invariant $\sim \beta$ is not generated, is valid for both approaches.  

The one loop contribution to spinor gravity (e.g. to the bosonic effective action) can
be written in terms of the matrix of second functional derivatives of the ``classical
action'' with respect to the
spinors in presence of a background $E^m_\mu,S^{(2))}\sim E{\cal D}$, as
\begin{eqnarray}\label{PB17}
\Gamma_{(1l)}&=&-\frac{1}{2}Tr~\ln(E{\cal D})~,\\
{\cal D}&=&\gamma^\mu\partial_\mu+\frac{1}{2E}\gamma^m
\partial_\mu(EE^\mu_m)=\gamma^\mu\hat{D}_\mu,\label{PB18}\\
\gamma^\mu&=&E^\mu_m\gamma^m. \label{PB16}
\end{eqnarray}
Here the classical action typically obtains after partial bosonization \cite{SG1} and we
have assumed here the form of eq. (\ref{F8a}) with $Z_a=0$. This holds for a large class
of actions of spinor gravity (far beyond the particular ansatz $S_E$ (\ref{7})) and a large
class of partial bosonizations. It simply reflects the absence of a classical spinor
coupling to an antisymmetric tensor. The coupling $Z_a$ can be incorporated easily, however.
Since it is typically generated by fluctuations it should be included in a nonperturbative
Schwinger-Dyson approach as discussed in the next section.
We call ${\cal D}$ the generalized Dirac operator
and observe the appearance of a ``covariant derivative''
\begin{equation}\label{PB19}
\hat{D}_\mu=\partial_\mu+\frac{1}{2E}E^m_\mu\partial_\nu(EE^\nu_m).
\end{equation}
(For Weyl spinors one should either multiply ${\cal D}$ by an appropriate projection operator
$(1+\bar{\gamma})/2$ or work within a reduced space of spinor indices, using $C\gamma^m$
instead of $\gamma^m$ since only $C\gamma^m$ acts in the reduced space.)
The contribution from the derivative acting on the vielbein can also be written in the form
\begin{equation}\label{33AX}
{\cal D}=\gamma^m(E^\mu_m\partial_\mu-\Omega_m)~,~\Omega_m=-\frac{1}{2E}\partial_\mu
(EE^\mu_m).
\end{equation}

It is instructive to compare the generalized Dirac
operator ${\cal D}$ with the corresponding operator
${\cal D}_E$ in Einstein-gravity. The latter is constructed from the Lorentz covariant
derivative $D_\mu$ which appears in the spinor kinetic term (Majorana spinors)
\begin{eqnarray}\label{PB20}
i\bar{\psi}\gamma^\mu D_\mu\psi&=&i\bar{\psi}\gamma^me^\mu_m
\left(\partial_\mu-\frac{1}{2}\omega_{\mu np}\Sigma^{np}\right)\psi
=i\bar{\psi}{\cal D}_E\psi\nonumber\\
&=&i\bar{\psi}\gamma^\mu\partial_\mu\psi-\frac{i}{4}\Omega_{[mnp]}
\bar{\psi}\gamma^{mnp}_{(3)}\psi.
\end{eqnarray}
Here $\gamma^{mnp}_{(3)}$ is the totally antisymmetrized product of three $\gamma$-matrices
$\gamma^{mnp}_{(3)}=\gamma^{[m}\gamma^n\gamma^{p]}$ and $\Omega_{[mnp]}$ corresponds
to the total antisymmetrization of
\begin{equation}\label{PB21}
\Omega_{mnp}=-\frac{1}{2}e^\mu_me^\nu_n(\partial_\mu e_{\nu p}-\partial_\nu e_{\mu p}).
\end{equation}
Replacing $e^m_\mu$ by $E^m_\mu$ one finds
\begin{equation}\label{PB22}
{\cal D}={\cal D}_E[E]+\frac{1}{4}\Omega_{[mnp]}[E]\gamma^{mnp}_{(3)}.
\end{equation}
For the fermionic one loop contribution the
only difference between spinor gravity and standard
gravity concerns the piece $\sim \Omega_{[mnp]}$! At this place we also note that a more
general ``classical action'' of the type (\ref{F8a}) with $Z_a\neq 0$ only modifies the
coupling of the second term in eq. (\ref{PB22}) by a factor
$\rho=(1-Z_a/Z_\psi)$ (cf. eq. (\ref{F16})).

Neglecting the piece $\sim\Omega_{[mnp]}$ the first contribution ${\cal D}_E[E^m_\mu]$ is
covariant with respect to both general coordinate and {\em local} Lorentz
transformations. Replacing ${\cal D}\rightarrow {\cal D}_E$ in the integral (\ref{PB17}) will
therefore lead to a one loop effective action $\Gamma_{1l}$ with these symmetries.
This is a gravitational effective action of the standard type. Expanded in the
number of derivatives one will find the curvature
scalar plus higher derivative invariants like $R^2,R_{\mu\nu}R^{\mu\nu}$ etc.
However, the additional piece $\sim\Omega_{[mnp]}[E^m_\mu]$ violates the local Lorentz
symmetry and only preserves a global Lorentz symmetry. We therefore expect the
appearance of new terms in the effective action which are invariant under global but
not local Lorentz-rotations. According to eq. (\ref{PB22}) all additional terms must
involve $\Omega_{[mnp]}$ or derivatives thereof. They vanish for $\Omega_{[mnp]}=0$.

We finally observe that the trace in eq. (\ref{PB17}) involves a trace over spinor indices as well as an
integration over space coordinates or, equivalently, a momentum integral in Fourier space.
As it stands, these integrations are highly divergent in the ultraviolet and the integral
(\ref{PB17}) needs a suitable regularization. This regularization should preserve the
invariance under general coordinate transformations. If possible, it should also preserve
the global Lorentz symmetry. However, there may be obstructions in the form of
``gravitational anomalies'' \cite{AW} for $d=6~mod~4$. At present it is not known if such
anomalies occur in spinor gravity. For the time being we neglect this possible complication
and assume global Lorentz symmetry of the effective action.

To gain some insight into the general structure of the one loop expression we may use the
nonlinear formulation. For this purpose we express the operator ${\cal D}$ as
\begin{equation}\label{F20}
{\cal D}=\tilde{H}\left({\cal D}_E[e]+\frac{i}{4}
K_{[mnp]}[e,H]\gamma^{mnp}_{(3)}\right)\tilde{H}^{-1}.
\end{equation}
In the one loop expression $\sim Tr \ln{\cal D}$ the factors $\tilde{H}$ and
$\tilde{H}^{-1}$ drop out. We conclude that all modifications of Einstein gravity
must involve the totally antisymmetric tensor $K_{[mnp]}$. In other words, the
dependence of $\Gamma_{1l}$ on $H$ appears only through $K_{[mnp]}$, i.e.
\begin{equation}\label{F21}
\Gamma_{1l}[e,H]=\Gamma_{1l}\big[e,K[e,H]\big].
\end{equation}
We observe that only the totally antisymmetric part
$K_{[mnp]}$ enters the one loop expression, whereas the contraction $K_m$ is absent.

In particular, in one loop order the additional invariant involving
two derivatives must be quadratic in $K_{[mnp]}$
\begin{eqnarray}\label{F22}
\Gamma_K&=&\frac{3\tilde{\tau}}{2}\int d^dxeK_{[mnp]}K^{[mnp]}\nonumber\\
&=&\frac{3\tilde{\tau}}{2}\int d^dxeU_{[\mu\nu\rho]}U^{[\mu\nu\rho]}.
\end{eqnarray}
Using the definitions (\ref{G2}) (\ref{S12}) this can be written in terms of $e,H$ or $E$
\begin{eqnarray}\label{F23}
\Gamma_K[e,H]&=&\frac{\tilde{\tau}}{2}\int d^dxe
\{\bar{D}_\mu e^m_\nu \bar{D}^\mu e^\nu_m\nonumber\\
&&-2\bar{D}_\mu e^m_\nu \bar{D}^\nu e^\mu_m\},\\
\Gamma_K[E]&=&\tilde{\tau}I_2.\label{F24}
\end{eqnarray}
Comparing the last expression with eq. (\ref{27NNA}) identifies
\begin{equation}\label{F25}
\tau=\tilde{\tau}/\mu\qquad,~\beta=0
\end{equation}
for one loop spinor gravity.

The one loop term can partly be found in the literature
\cite{Ak}, \cite{DS}, \cite{Tor} or combined from it. For the completeness of this paper
we find it useful, nevertheless, to describe the most important aspects explicitely in
our formulation. From eq. (\ref{F16}) we conclude that we can write the
one loop expression as a
standard gravitational one loop correction for the effective action of the vielbein
$e^m_\mu$, plus a torsion term
\begin{equation}\label{LL1a}
\Gamma_{1l}[e,K]=-\frac{1}{2}Tr\ln\left\{e{\cal D}_E[e]+
\frac{\rho}{4}eK_{[mnp]}\gamma^{mnp}_{(3)}\right\}.
\end{equation}
Here we use
\begin{equation}\label{LL2a}
{\cal D}_E[e]=\gamma^me^\mu_m\partial_\mu-\frac{1}{4}\Omega_{[mnp]}[e]
\gamma^{mnp}_{(3)}
\end{equation}
and $\rho=(1-Z_a/Z_\psi)$. For $K_{[mnp]}=0$ one
recovers the one loop expression for standard gravity.
In lowest order in the derivative expansion this produces a term proportional to the
curvature scalar in the effective action.

The new ingredient in spinor gravity involves the contribution of generalized torsion
associated to the tensor $K_{[mnp]}$. For any regularization that preserves the
diffeomorphism symmetry the new terms involving $K$ must be invariant under general
coordinate transformations. The same holds for the local Lorentz transformations. We may
expand in powers of $K$. There is no possible term
linear in $K$ since no totally antisymmetric tensor can be constructed from $e^m_\mu$.
The term quadratic in $K$ reads (\ref{F22})
\begin{eqnarray}\label{LL4a}
\Gamma_K&=&\frac{\rho^2}{64}Tr\{{\cal D}^{-1}_EA_K{\cal D}^{-1}_EA_K\}
=\tilde{\tau}\int d^dxe K_{[mnp]}K^{[mnp]}\nonumber\\
A_K&=&K_{[mnp]}\gamma^{mnp}_{(3)}.
\end{eqnarray}
One may compute $\tilde{\tau}$ by evaluating $\Gamma_K$ for constant ($x$-independent) values
of $K$ and $e^m_\mu=\delta^m_\mu$. This allows us to perform a calculation in flat space
and to use a Fourier basis. With ${\cal D}_E\rightarrow i\gamma^\mu p_\mu~,~
{\cal D}^{-1}_E\rightarrow-i\gamma^\mu p_\mu/p^2~,~p^2=p_\mu p^\mu$ one finds
\begin{eqnarray}\label{LL5a}
\Gamma_K&=&-\frac{\rho^2}{64}\Omega_d\int \frac{d^dp}{(2\pi)^d}\frac{p_\mu p_\nu}{p^4}
K_{[\rho_1\rho_2\rho_3]}K^{[\sigma_1\sigma_2\sigma_3]}\nonumber\\
&&tr\{\gamma^\mu\gamma^{\rho_1\rho_2\rho_3}_{(3)}\gamma^\nu\gamma_{(3)\sigma_1
\sigma_2\sigma_3}\}.
\end{eqnarray}
Here $\Omega_d$ denotes the $d$-dimensional volume and the last trace $tr$ is only
in spinor space.

Obviously, the momentum integral is ultraviolet divergent and needs a regularization. We
may evaluate it in a euclidean setting and use rotation symmetry
\begin{equation}\label{LL6a}
\int d^dpf(p^2)p_\mu p^\nu=\frac{1}{d}\int d^dpf(p)p^2\delta^\nu_\mu.
\end{equation}
We may also employ
\begin{eqnarray}\label{LL7a}
tr \gamma^\mu&\gamma^{\rho_1\rho_2\rho_3}_{(3)}&\gamma_\mu
\gamma_{(3)\sigma_1\sigma_2\sigma_3}\nonumber\\
&=&-(d-6)tr\gamma_{(3)}^{\ \ \rho_1\rho_2\rho_3}
\gamma_{(3)\sigma_1\sigma_2\sigma_3} \nonumber\\
&=&-(d-6)q_d~\delta^{[\rho_1\rho_2\rho_3]}_{\sigma_1\sigma_2\sigma_3}
\end{eqnarray}
such that $(\rho=1)$
\begin{equation}\label{LL8a}
\tilde{\tau}=\frac{(d-6)q_d}{96d}\int
\frac{d^dp}{(2\pi)^d}\frac{1}{p^2}.
\end{equation}
With $(x=p^2)$
\begin{equation}\label{LL9a}
\int\frac{d^dp}{(2\pi)^d}=2v_d\int dxx^{d/2-1}~,~v^{-1}_d=
2^{d+1}\pi^{d/2}\Gamma\left(\frac{d}{2}\right)
\end{equation}
and
\begin{equation}\label{LL10a}
\int dxx^{\kappa-1}=\frac{1}{\kappa}A_{2\kappa}\Lambda^{2\kappa}
\end{equation}
we put the effects of the particular regularization scheme into the dimensionless
coefficient $A_{2\kappa}(A_{2\kappa}=1$ for a sharp momentum cutoff). We see that
$\tau$ depends strongly on the regularization scale (or $UV$ cutoff) $\Lambda$
\begin{equation}\label{LL11a}
\tilde{\tau}=\frac{v_d(d-6)q_d}{24d(d-2)}A_{d-2}
\Lambda^{d-2}.
\end{equation}
For $d>6$ the coefficient $\tilde{\tau}$ is positive.

Already at this rather primitive level one sees that the one loop correction
induces a gravitational effective action containing the curvature scalar as well as a new
torsion invariant involving the square of the totally antisymmetric tensor $K_{[mnp]}$.
This corresponds to the invariant $I_2$ with coupling $\tau$
in sect. \ref{globallorentzsymmetry}
whereas $\beta$ vanishes
in the one loop approximation.

Obviously the dependence on the regularization is disturbing, in particular since a simple
momentum cutoff is not compatible with diffeomorphism symmetry. This does not permit
us to compute the overall size of the one loop expression in a regularization
independent way. However, at least the ratio between the
coefficients of the torsion and curvature invariants is independent of the
particular regularization, provided that diffeomorphism symmetry is preserved. Employing
standard techniques \cite{Tor} one finds for $d=4$ \cite{HW1},
\begin{equation}\label{ZX}
\tau=3\left(1-\frac{Z_a}{Z_\psi}\right)^2
\end{equation}

\section{Schwinger-Dyson equation}
\label{schwingerdysonequation}
In the preceeding section we have sketched the one loop contribution of spinor
fluctuations in a background field $E^m_\mu$. This computation is also a central piece in a
nonperturbative approach based on the Schwinger-Dyson equations \cite{SDE}. Following the
treatment of \cite{BEA},\cite{JW} we use a general notation for the spinors
$\psi_\alpha$ where $\alpha$ collects all spinor indices, i.e. internal indices as well as
momentum labels. We further assume here for simplicity
that the classical action is a polynomial of $n$
spinor fields
\begin{equation}\label{SD1}
S=\frac{1}{n!}\lambda_{\alpha_1\dots\alpha_n}
\psi_{\alpha_1}\dots\psi_{\alpha_n}
\end{equation}
with $\lambda_{\alpha_1\dots\alpha_n}$ totally antisymmetric in all $n$ indices.
The  fermionic effective action can be written as an implicit functional integral
\begin{equation}\label{SD2}
\Gamma_F[\psi]=-\ln\int{\cal D}\tilde{\psi}\exp
\big\{-S[\psi+\tilde{\psi}]+\eta_\alpha\tilde{\psi}_\alpha\big\}
\end{equation}
where
\begin{equation}\label{SD3}
\eta_\alpha[\psi]=-\frac{\delta\Gamma_F}{\delta\psi_\alpha}.
\end{equation}
Taking a derivative with respect to $\psi_\alpha$ one obtains
\footnote{Note $\langle\tilde{\psi}_\alpha\rangle=0$.} the exact identity
$(\psi'=\psi+\tilde{\psi})$
\begin{eqnarray}\label{SD4}
\frac{\delta\Gamma_F}{\delta\psi_\alpha}&=&
\left\langle\frac{\delta S}{\delta\tilde{\psi}_\alpha}
[\psi+\tilde{\psi}]\right\rangle\nonumber\\
&=&\frac{1}{(n-1)!}\lambda_{\alpha\alpha_2\dots\alpha_n}
\langle\psi'_{\alpha_2}\dots\psi'_{\alpha_n}\rangle.
\end{eqnarray}

We next express the r.h.s of eq. (\ref{SD4}) in terms of the fermion propagator
\begin{equation}\label{SD5}
G_{\alpha\beta}=\langle\psi'_\alpha\psi'_\beta\rangle
\end{equation}
as
\begin{equation}\label{SD6}
\frac{\delta\Gamma_F}{\delta\psi_\alpha}=\frac{1}{F_n}
\lambda_{\alpha\beta\alpha_3\dots\alpha_n}\psi_\beta
G_{\alpha_3\alpha_4}\dots G_{\alpha_{n-1}\alpha_n}+X_\alpha+Y_{\alpha\beta}\psi_\beta
\end{equation}
with
\begin{equation}\label{SD7}
F_n=(n-2)(n-4)\dots 2
\end{equation}
Here $X_\alpha$ contains terms with more than one power of $\psi$ and $Y_{\alpha\beta}$
involves connected Greens functions for more than two spinors. Taking one more derivative
of eq. (\ref{SD6}) and evaluating it at $\psi=0$ yields the exact Schwinger-Dyson
equation (SDE)
\begin{eqnarray}\label{SD8}
\left(\Gamma^{(2)}_F\right)_{\alpha\beta}&=&
\frac{\partial^2\Gamma}{\partial\psi_\beta\partial\psi_\alpha}
=(G^{-1})_{\alpha\beta}\\
&=&\frac{1}{F_n}\lambda_{\alpha\beta\alpha_3\dots\alpha_n}
G_{\alpha_3\alpha_4}\dots G_{\alpha_{n-1}\alpha_n}
+Y_{\alpha\beta}.\nonumber
\end{eqnarray}
The lowest order SDE neglects the contribution $Y_{\alpha\beta}$. It is therefore a
closed equation for the propagator  $G_{\alpha\beta}$.

Let us define an effective action for the propagator $G_{\alpha\beta}$
\begin{eqnarray}\label{SD9}
\Gamma[G]&=&\frac{1}{nF_n}\lambda_{\alpha_1\dots\alpha_n}G_{\alpha_1\alpha_2}\dots
G_{\alpha_{n-1}\alpha_n}
+\frac{1}{2}Tr\ln G\nonumber\\
&=&\Gamma_{(cl)}+\Gamma_{(1)}
\end{eqnarray}
The functional variation of $\Gamma[G]$ with respect to $G_{\alpha\beta}$ yields
\footnote{In our notation $\partial G_{\alpha\beta}/\partial G_{\gamma\delta}=
\delta_{\alpha\gamma}\delta_{\beta\delta}-\delta_{\alpha\delta}\delta_{\beta\gamma}$.}
precisely eq. (\ref{SD8}) (for $Y_{\alpha\beta}=0$). We can therefore interprete
$\Gamma[G]$ as the effective action for the bosonic spinor bilinears $G_{\alpha\beta}$.
The field equations derived from this bosonic effective action correspond precisely
to the lowest order SDE. We may call the first term on the r.h.s. of eq. (\ref{SD7})
the classical bosonic action ($n=2d$)
\begin{eqnarray}\label{SD10}
\Gamma_{(cl)}[G]&=&
\frac{1}{nF_n}\int dx_1\dots dx_ddy_1\dots dy_d \nonumber\\
&&\lambda_{a_1\dots a_d b_1\dots b_d}
(x_1\dots x_d,y_1\dots y_d)\nonumber\\
&&G_{a_1b_1}(x_1,y_1)\dots G_{a_db_d}
(x_d,y_d)
\end{eqnarray}
and the second term the one loop term. Here we have used a double index notation
$\alpha=(a,x)$ where $x=x^\mu$ is a ``coordinate index'' indicating the location in space
and $a$ collects all internal indices, i.e. the spinor indices of the irreducible
representations of the Lorentz group and indices counting the possibly different
representations (e.g. $\psi$ and $\bar{\psi}$ in case they are not related by a
Majorana condition). We recall that the antisymmetrisation of $\lambda$ concerns the
permutations of all index pairs $(x_i,a_i)\leftrightarrow(x_j,a_j)$.

We need the coupling $\lambda$ corresponding to
the classical action $S_E=\alpha\int d^d\tilde{E}$
(\ref{7}). Comparison with (\ref{SD1}) yields
\begin{eqnarray}\label{SD11}
\lambda_{a_1\dots a_db_1\dots b_d}
(x_1\dots x_d,y_1\dots y_d)=
\left(-\frac{i}{2}\right)^d\frac{\alpha}{d!}\nonumber\\
P_A\Big\{\int d^dz\delta(x_1-z)\dots\delta(x_d-z)\delta(y_1-z)\dots \nonumber\\
\delta(y_d-z)\epsilon^{\mu_1\dots\mu_d}\epsilon_{m_1\dots m_d}\nonumber\\
\left(\frac{\partial}{\partial x_1^{\mu_1}}-\frac{\partial}{\partial y_1^{\mu_1}}\right)
\dots\left(\frac{\partial}{\partial x^{\mu_d}_d}-\frac{\partial}{\partial y^{\mu_d}_d}\right)
\nonumber\\
\hat{\gamma}^{m_1}_{a_1b_1}\dots\hat{\gamma}^{m_d}_{a_db_d}\Big\}.
\end{eqnarray}
Here $\hat{\gamma}^m_{ab}=(\hat{C}\gamma^m)_{ab}$ contains the charge conjugation matrix for
Majorana spinors $(\hat{C}=C)$ or an appropriate matrix in the $\bar{\psi}-\psi$ space
for Dirac spinors. For Weyl spinors it involves an additional projector $(1+\bar{\gamma})/2$.
We consider here symmetric matrices $\hat{\gamma}^m$,
$\hat{\gamma}^m_{ab}=\hat{\gamma}^m_{ba}$, in accordance with
the antisymmetry under the exchange
$(x_1a_1)\leftrightarrow(y_1b_1)$ etc.. The operator $P_A$ denotes the sum over all
permutations \footnote{The factor $1/n!$ appearing in the total antisymmetrisation is
already included in the definition (\ref{SD1}) of $\lambda$.} of index pairs with a
minus sign for odd permutations.

We want to evaluate $\Gamma$ in terms of the vielbein
\begin{eqnarray}\label{SD12}
E^m_\mu(x)&=&-\frac{i}{2}\langle\partial_\mu\psi'(x)\hat{\gamma}^m\psi'(x)
-\psi'(x)\hat{\gamma}^m\partial_\mu\psi'(x)\rangle\nonumber\\
&=&-\frac{i}{2}\left(\frac{\partial}{\partial x^\mu}-\frac{\partial}{\partial y^\mu}\right)
G_{ab}(x,y)_{|x=y}\hat{\gamma}^m_{ab}.
\end{eqnarray}
Variation of $\Gamma[E^m_\mu]$ with respect to $E^m_\mu$ yields a nonperturbative
approximation to the geometrical field equations of spinor gravity. This approximation
not only neglects the higher order terms $(Y_{\alpha\beta})$ but also the possible
impact of expectation values of fermion bilinears other than the vielbein.

One immediately realizes that $\Gamma_{(1)}$ has precisely the structure of the one loop
contribution $\Gamma_{(1l)}$ (\ref{PB17}) in the preceeding section. The only
difference is that $S^{(2)}=E{\cal D}$ is now replaced by the full inverse propagator
$G^{-1}=\Gamma^{(2)}_F$. At this point we can make use of the general discussion of
sect. \ref{gravitationalspinor} in order to parameterize the form of the inverse fermion
propagator in presence of background field $E^m_\mu$. In consequence, $\Gamma_{(1)}$
will now depend on unknown couplings like $Z_\psi$ and $Z_a$ (cf. eq. (\ref{F8a})). These
couplings have to be determined self-consistently by the solution of the Schwinger-Dyson
equation (\ref{SD8}) for $G_{ab}(x,y)$ with $x\neq y$. For a local classical action
(\ref{SD11}) and $Y_{\alpha\beta}=0$ the r.h.s. of eq. (\ref{SD8}) only consists of a
local term. In turn, this contains polynomials in the vielbein of order $n/2-1$.
Typically, it is also linear in the fermion momentum (from an ``unsaturated derivative''
corresponding to the indices $\alpha\beta$ in eq. (\ref{SD7})) and has therefore the
structure discussed in sect. \ref{gravitationalspinor}. (Care has to be taken for the
permutations in $P_A$.) Once $Z_a/Z_\psi$ has been determined one can take over the
discussion of sect. {\ref{oneloopspinorgravity}. This results in a contribution to the effective action where the terms involving two derivatives are given by eq. (\ref{AA2}), with $\tau$ specified by eq. (\ref{ZX}) and $\beta=0$.

We also need the classical contribution  $\Gamma_{(cl)}$. This obviously depends on the
choice of the action for spinor gravity. We demonstrate its computation for the choice
$S_E$ (\ref{7}). Even in this case the classical action is not simply
$\alpha\int d^dxE(x)$ since one has to take care of the permutations $P_A$.
For a computation of $\Gamma_{(cl)}$ we have to classify how the various combinations
of $\hat{\gamma}^m_{\hat{a}\hat{b}}$ appearing in the permutation sum are contracted
with $G_{ab}$. The first contribution are all terms where for each $\hat{\gamma}^m$ both
its indices are contracted with a suitable $G_{ab}$. There are
$2d\cdot(2d-2)\dots 2=nF_n$ such terms in the sum $P_A$. The contributions of these
terms to $\Gamma_{(cl)}$ is therefore precisely $(E=\det E^m_\mu)$
\begin{eqnarray}\label{SD13}
\Gamma^{(1)}_{(cl)}&=&\frac{\alpha}{d!}\int d^dx\epsilon^{\mu_1\dots \mu_d}
\epsilon_{m_1\dots m_d}
E^{m_1}_{\mu_1}(x)\dots E^{m_d}_{\mu_d}(x)\nonumber\\
&=&\alpha\int d^dxE(x).
\end{eqnarray}

These are, however, not all possible contractions. Let us next look at the term where
contractions of the type $\hat{\gamma}^m_{ab}G_{ab}$ occur only for $d-2$ of the
$\hat{\gamma}^m$-matrices, and not for the two others. There are again $nF_n$ such
terms. They lead to a second contribution
\begin{eqnarray}\label{SD14}
\Gamma^{(2)}_{(cl)}&=&\frac{\alpha}{d!}\int d^dx\epsilon^{\mu_1\dots\mu_d}
\epsilon_{m_1\dots m_d}E^{m_3}_{\mu_3}(x)\dots \nonumber\\
&&E^{md}_{\mu d}(x)
L^{m_1m_2}_{\mu_1\mu_2}(x),\\
L^{m_1m_2}_{\mu_1\mu_2}(x)&=&-\frac{1}{8}
\left(\frac{\partial}{\partial x^{\mu_1}_1}-\frac{\partial}{\partial y^{\mu_1}_2}\right)
\left(\frac{\partial}{\partial x^{\mu_2}_2}-\frac{\partial}{\partial y^{\mu_2}_1}\right)\nonumber\\
&&\left(\hat{\gamma}^{m_1}_{a_1b_2}\hat{\gamma}^{m_2}_{a_2b_1}
-\hat{\gamma}^{m_2}_{a_1b_2}\hat{\gamma}^{m_1}_{a_2b_1}\right)\label{SD15}
\\
&&G_{a_1b_1}(x_1,y_1)G_{a_2b_2}(x_2,y_2)_{|x_1=x_2=y_1=y_2=x}\nonumber
\end{eqnarray}
In order to proceed further we need a generalized Fierz identity
\begin{equation}\label{SD16}
\gamma^{m_1}_{a_1b_2}\gamma^{m_2}_{a_2b_2}=
c\gamma^{m_1}_{a_1b_1}\gamma^{m_2}_{a_2b_2}+\dots
\end{equation}
where the dots denote products of other elements of the Clifford algebra
$\Gamma_{a_1b_1}\Gamma'_{a_2b_2}$ like $\Gamma=\gamma^{[mnp]}$ etc.. The first
term $\sim c$ could change the coefficient $\alpha$ in eq. (\ref{SD13}) whereas the other
terms do not contribute in an approximation where we neglect possible expectation values
of  local fermion bilinears except the vielbein. A similar procedure holds for the other
possible contractions.

We know from symmetry arguments that the only local invariant involving the vielbein
without derivatives has the structure of eq. (\ref{SD13}) such that the additional
contractions may only modify the effective coefficient $\alpha$. In principle, the
contractions of the type of eq. (\ref{SD14}) and beyond could produce invariants involving
derivatives of the vielbein. However, such invariants would have to be polynomial in the
vielbein. The existence of such polynomials is severely restricted \cite{SG1} and they do
not occur in a parity invariant setting or for $d\neq 4$. If we only consider nonvanishing
$E^m_\mu$ (and put all other expectation values of local fermion bilinears to zero) the
symmetries alone imply that the classical action must be $\sim\int d^dxE$. This covers
arbitrary invariant actions of spinor gravity, including all those discussed in sect.
\ref{spinoraction}.

For a particular setting where the spinor action contains precisely $2d$ spinors we also
observe that $d$ derivatives have to match precisely $d$
fermion bilinears. If more
than one derivative acts on a given bilinear (for example $\partial_\mu E^m_\nu$
involves two derivatives on the fermions), another one must have zero derivatives, which
is not allowed if the only nonvanishing bilinears are real $E^m_\mu$. Obviously, this
discussion could change if nonvanishing expectation values of non-derivative bilinears
like $\langle \bar{\psi}\gamma^{mnp}\psi\rangle$ or $\langle \bar{\psi}\gamma^m\psi\rangle$
occur. Finally, we note that the contribution $\sim c$ in $L^{m_1m_2}_{\mu_1\mu_2}$ actually
vanishes if $\langle\bar{\psi}\gamma^m\psi\rangle=0$.

We conclude that the Schwinger-Dyson equations yield in lowest order a result very similar
to the one loop approximation, except that the numerical coefficients are possibly modified and
have to be determined self-consistently. In particular, they also imply $\beta=0$ in eq.
(\ref{AA2}). These statements hold independently of the form of the action of spinor
gravity as long as diffeomorphism and global Lorentz symmetry is preserved and the spinor
kinetic term has the form (\ref{F8a}).

\section{Conclusions}
\label{conclusions}

Spinor gravity is an interesting candidate for a quantum theory of gravity. We have demonstrated the existence of a large but finite number of invariants that are polynomials in the fermionic Grassmann variables and invariant under the symmetries of diffeomorphisms and global Lorentz rotations. The central missing piece for the completion of a well defined functional integral defining quantum gravity remains the formulation of a regularized functional measure consistent with these symmetries. A further open question concerns the role of local versus global Lorentz-symmetry: is it necessary to require the invariance of the action under local Lorentz-symmetry?

In order to make progress on this last point we have investigated the models with global, but not local Lorentz symmetry. We have concentrated here on the general geometrical features that are independent of the precise form of the classical action. Using nonlinear fields it becomes apparent that the difference to Einstein's gravity (with local Lorentz symmetry) consists in the presence of new massless ``Goldstone-type'' fields. This leads to a generalization of the geometrical concepts in gravity, like modified covariant derivatives or torsion.

The structure of the effects of quantum fluctuations in spinor gravity becomes apparent by an investigation of the solutions to the Schwinger-Dyson equation (SDE) in lowest order. They result in an effective gravitational action (and corresponding effective gravitational field equations) which contains the usual curvature invariants for the metric plus additional kinetic terms for the ``Goldstone degrees of freedom''. To lowest order the (non-perturbative) expansion of the SDE yields a new invariant with a particular structure. Precisely this structure is compatible \cite{SG1} with all present observations despite the lack of local Lorentz symmetry.

\vspace{1cm}
\noindent
{\em Acknowledgment}

The author would like to thank to thank A. Hebecker for collaboration on
several key points, as summarized in \cite{HW1}.

\section* {Appendix A: Local Lorentz transformations}
\renewcommand{\theequation}{A.\arabic{equation}}
\setcounter{equation}{0}
In this appendix we study the action of {\em local} Lorentz transformations on the
invariants with respect to the global Lorentz-transformations that we have discussed
in section \ref{spinoraction}.
Our motivation is to find out if there exists a polynomial action for
spinors that is also invariant under local Lorentz-transformations. This would reduce the
gravitational degrees of freedom and bring spinor gravity much closer to Einstein gravity. A study of the local Lorentz-transformation properties of the various bilinears is also useful for a discussion of the properties of the loop expansion or solutions to SDE.

With respect to the local Lorentz rotations the transformation
properties also involve an inhomogeneous piece
\begin{equation}\label{T1}
\delta_{\cal L}\tilde{E}^m_\mu=\epsilon^m_{\ n}\tilde{E}^n_\mu
+\frac{i}{4}\bar{\psi}\gamma^{mnp}_{(3)}\psi\partial_\mu\epsilon_{np}
+\frac{i}{2}\bar{\psi}\gamma^n\psi\partial^\mu\epsilon^m_{\ n}
\end{equation}
with
\begin{eqnarray}\label{T2}
\gamma^{mnp}_{(3)}&=&\gamma^{[m}\gamma^n\gamma^{p]}\nonumber\\
&=&\frac{1}{6}(\gamma^m\gamma^n\gamma^p-\gamma^m\gamma^p\gamma^n
+\gamma^n\gamma^p\gamma^m\nonumber\\
&&- \gamma^n\gamma^m\gamma^p+
\gamma^p\gamma^m\gamma^n-\gamma^p\gamma^n\gamma^m).
\end{eqnarray}
We recall that for Majorana spinors in $d=2,3,9~ mod~8$ the
antisymmetry under the exchange of Grassmann
variables implies $\bar{\psi}\gamma^n\psi\equiv 0$ \cite{CWMS}. The part
$\sim\bar{\psi}\gamma^{mnp}_{(3)}\psi$ remains, however. At this point we observe that for
Majorana-Weyl spinors in $d=2~ mod~ 8$ the Weyl constraint implies $\bar{\psi}\Gamma^k\psi=0$
if $\Gamma^k$ is a product of an even number of $\gamma$-matrices, and the required
antisymmetry of $C\Gamma^k$ admits nonvanishing nonderivative
bilinears only if $\Gamma^k$ is the totally antisymmetric product of
$k=3~ mod~ 4~\gamma$-matrices \cite{CWMS}.

As an example, we will investigate the {\em local} Lorentz transformations
for a MW-spinor in $d=10$. In terms of $\tilde{Q}$ (\ref{A4})
we can write the inhomogeneous part of the Lorentz transformations (\ref{T1}) as
\begin{equation} \label{A5}
\delta_{in}\tilde{E}^m_\mu=\tilde{Q}^{mnp}\partial_\mu\epsilon_{np}.
\end{equation}
(We do not list the trivial homogeneous part which simply reflects the tensor
structure.) Exploiting the vanishing of $\bar{\psi}\gamma^m\psi$ and
$\bar{\psi}\gamma^{mnpqs}_{(5)}\psi$ for Majorana-Weyl spinors we find the
transformation properties of $\tilde{F}$ and $\tilde{G}$
\begin{eqnarray}\label{A6}
\delta_{in}\tilde{F}_\mu^{\ m_1m_2m_3}&=&-\frac{i}{2}\bar{\psi}\gamma^{m_1m_2m_3}_{(3)}
\Sigma^{np}\psi\partial_\mu\epsilon_{np}\\
&=&2(\eta^{m_1n}\tilde Q^{pm_2m_3}+\eta^{m_2n}\tilde Q^{m_1pm_3}\nonumber\\
&&+\eta^{m_3n}\tilde Q^{m_1m_2p})\partial_\mu\epsilon_{np}
\end{eqnarray}
and
\begin{eqnarray}\label{A7}
\delta_{in}\tilde{G}_\mu^{\ m_1m_2m_3m_4m_5}&=&-\frac{i}{2}\bar{\psi}
\gamma^{m_1m_2m_3m_4m_5}_{(5)}{\Sigma}^{np}\psi
\partial_\mu\epsilon_{np}\nonumber\\
=&\{\frac{\alpha}{6}&\epsilon^{m_1\dots m_5nps_1s_2s_3}\tilde Q_{s_1s_2s_3}\\
&-&20 A_5[\eta^{m_1n}\eta^{m_2p}\tilde Q^{m_3m_4m_5}]\}
\partial_\mu\epsilon_{np}.\nonumber
\end{eqnarray}
Here we have used the Weyl constraint $\bar{\gamma}\psi=\psi$ with
\begin{eqnarray}\label{A8}
&&\bar{\gamma}=\eta\epsilon_{m_1\dots m_d}\gamma^{m_1\dots m_d}~,~  \nonumber\\
&&\gamma^{m_1\dots m_7}_{(7)}\bar{\gamma}=\frac{\hat{\alpha}}{6}
\epsilon^{m_1\dots m_7n_1n_2n_3}
\gamma_{(3)n_1n_2n_3}
\end{eqnarray}
and $\eta=1,\hat{\alpha}=1$ for Minkowski signature $\eta_{mn}=(-1,+1,+1\dots)$.
The operation
$A_5$ denotes total antisymmetrization over the indices $(m_1,m_2,m_3,m_4,m_5)$ such
that the second term in eq. (\ref{A7}) contains 20 independent combinations,
$20 A_5[\eta^{m_1n}\eta^{m_2p}\tilde Q^{m_3m_4m_5}]=
(\eta^{m_1n}\eta^{m_2p}-\eta^{m_1p}\eta^{m_2n})\tilde Q^{m_3m_4m_5}-
(m_2\leftrightarrow m_3)
+(m_2\leftrightarrow m_4)-\cdots$.

Similarly, the
inhomogeneous part of the local Lorentz transformation of the tensors
$\tilde{S}, \tilde{T}$ (\ref{B9A}), (\ref{B2}) reads
\begin{eqnarray}\label{B11}
\delta_{in}\tilde{S}_{\mu_1\mu_2}^{\ \ \ \ m}&=&\partial_{\mu_1}\tilde Q^{mnp}
\partial_{\mu_2}\epsilon_{np}\nonumber\\
&&+\partial_{\mu_1}\epsilon^m_{\ n}
\tilde{E}^n_{\mu_2}-(\mu_1\leftrightarrow\mu_2)
\end{eqnarray}
and
\begin{eqnarray} \label{B12}
&&\delta_{in}\tilde{T}_{\mu_1\mu_2}^{\ \ \ \ m_1\dots m_5}\nonumber\\
&=&\frac{\alpha}{6}\epsilon^{m_1\dots m_5nps_1s_2s_3}
\partial_{\mu_1}\tilde Q_{s_1s_2s_3}
\partial_{\mu_2}\epsilon_{np}\nonumber\\
&&-20A_5[\partial_{\mu_1}\tilde Q^{m_3m_4m_5}\partial_{\mu_2}
\epsilon^{m_1m_2}]\nonumber\\
&&+\partial_{\mu_1}\epsilon^{m_1}_{\ n}\tilde{G}^{nm_2m_3m_4m_5}\nonumber\\
&&+\partial_{\mu_1}\epsilon^{m_2}_{\ n}\tilde{G}^{m_1nm_3m_4m_5}
+\cdots \nonumber\\
&&+\partial_{\mu_1}\epsilon^{m_5}_{\ n}
\tilde{G}^{m_1m_2m_3m_4n}-(\mu_1\leftrightarrow\mu_2).
\end{eqnarray}

We can combine these transformation properties in order to find the change of global
Lorentz invariants under local Lorentz transformations. The global invariance ensures that
the homogeneous part of the transformation vanishes, such that we need only the
contribution from $\delta_{in}$. The invariant $I_E$ (\ref{7}) which involves only powers of
$\tilde{E}^m_\mu$ transforms as \footnote{
For the case where $\tilde{E}=0$ and no inverse vielbein $\tilde{E}^\mu_m$ exists one
should consider $\tilde{E}\tilde{E}^\mu_m$ as a shorthand for the polynomial defined by
eq. (\ref{S5}).}
\begin{equation}\label{B18}
\delta_{{\cal L}}I_E=\delta_{{\cal L}}\int d^dx\tilde{E}=\int d^dx \tilde{E}\tilde{E}^\mu_m
\tilde Q^{mnp}\partial_\mu\epsilon_{np}.
\end{equation}
Clearly, this is not invariant for general functions $\epsilon_{np}(x)$. Nevertheless,
some residual invariance remains for a special class of functions $\epsilon_{np}(x)$ which
is defined in dependence on the value of the spinor bilinear $\tilde{E}^m_\mu$. If
we define $\tau_{mnp}=\tilde{E}\tilde{E}^\mu_m\partial_\mu\epsilon_{np}$ one observes
that the right hand side vanishes if the totally antisymmetric part of $\tau$ vanishes
$\tau_{[mnp]}=0$.
This is simply a consequence of the contraction with the totally antisymmetric quantity
$Q^{mnp}$. Since the condition $\tau_{[mnp]}=0$ involves the spinor fields
in a nonlinear fashion the local transformations which preserve the relation
$\tau_{[mnp]}(x)=0$ form a group of nonlinear field transformations. One may speculate that
such a nonlinear symmetry is responsible for the absence of the contraction $K_m$ in
the one loop expression (\ref{LL4a}) and therefore for $\beta=0$.

Next we consider the invariant
\begin{eqnarray}\label{B19}
I_{\Omega K}&=&\int d^dx\tilde{E}\tilde{\Omega}_{[mnp]}\tilde Q^{mnp}\\
&=&-\frac{1}{4(d-2)!}\int d^dx\epsilon^{\mu_1\dots\mu_d}\epsilon_{m_1\dots m_d} \nonumber\\
&& \qquad \tilde{E}^{m_3}_{\mu_3}\dots\tilde{E}^{m_d}_{\mu_d}
\tilde{S}_{\mu_1\mu_2 p}\tilde Q^{m_1m_2p}\nonumber
\end{eqnarray}
where
\begin{equation}\label{B20}
\tilde{E}\tilde{\Omega}_{mnp}=\tilde{E}\tilde{E}^\mu_m\tilde{E}^\nu_n
\tilde{\Omega}_{\mu\nu p}=-\frac{1}{2}
\tilde{E}\tilde{E}^\mu_m\tilde{E}^\nu_n
\tilde{S}_{\mu\nu p}.
\end{equation}
This is motivated by the observation that the combination $I_E-I_{\Omega K}$ bares
some resemblance with a covariant derivative with spin connection
$\tilde{\Omega}$.
With eqs. (\ref{A5}), (\ref{B11}) and $\delta_{in}\tilde Q^{mnp}=0$ one obtains
\begin{eqnarray}\label{B21}
\delta_{{\cal L}}I_{\Omega K}&=&-\frac{1}{4(d-2)!}\int d^dx\epsilon^{\mu_1\dots\mu_d}
\epsilon_{m_1\dots m_d}\tilde Q^{m_1m_2p}\nonumber\\
&&\Big\{(d-2)\tilde{E}^{m_4}_{\mu_4}\dots \tilde{E}^{m_d}_{\mu_d}\tilde Q^{m_3st}
\partial_{\mu_3}\epsilon_{st}\tilde{S}_{\mu_1\mu_2 p}\nonumber\\
&&+2\tilde{E}^{m_3}_{\mu_3}\dots\tilde{E}^{m_d}_{\mu_d}
(\partial_{\mu_1}\tilde Q_p^{\ st}\partial_{\mu_2}\epsilon_{st}+
\partial_{\mu_1}\epsilon_{ps}\tilde{E}^s_{\mu_2})\Big\}\nonumber\\
&=&\int d^dx\tilde{E}\tilde Q^{mnp}\Big\{\tilde{E}^\mu_m\partial_\mu\epsilon_{np}\nonumber\\
&&-\frac{3}{2}\tilde{E}^\mu_{[m}\tilde{E}^\nu_n\tilde{E}^\rho_{q]}\tilde{S}_{\mu\nu p}
\tilde Q^{qst}\partial_\rho\epsilon_{st}\nonumber\\
&&-\tilde{E}^\mu_m\tilde{E}^\nu_n\partial_\mu\tilde{Q}_p^{\ st}\partial_\nu\epsilon_{st}\Big\}\nonumber\\
&=&\int d^dx\tilde{E}\tilde Q^{mnp}
\Big\{\partial_m\epsilon_{np}-\partial_m \tilde Q_p^{\ st}\partial_n\epsilon_{st}\nonumber\\
&&+\tilde Q^{qst}\tilde{S}_{mqp}\partial_n\epsilon_{st}
-\frac{1}{2}\tilde{Q}^{qst}\tilde{S}_{mnp}\partial_q\epsilon_{st}\Big\}.
\end{eqnarray}
We observe that $\partial\epsilon$ now multiplies new structures $\tilde{Q}\partial\tilde{Q}$
and $\tilde{Q}\tilde{Q}\tilde{S}$. These new types of structures can actually be related to
each other.
Using
\begin{equation}\label{B22}
\partial_\mu \tilde{E}=\tilde{E}\tilde{E}^\nu_n\partial_\mu
\tilde{E}^n_\nu~,~\partial_\mu \tilde{E}^\nu_n=-\tilde{E}^\nu_p\tilde{E}^\rho_n
\partial_\mu \tilde{E}^p_\rho~,
\end{equation}
one finds by partial integration
\begin{eqnarray}\label{B23}
&&\int d^dx\tilde{E}\tilde{E}^\mu_m\tilde{E}^\nu_n~
\tilde Q^{mnp}\partial_\mu \tilde Q_p^{\ st}\partial_\nu\epsilon_{st}\nonumber\\
&=&-\int d^dx\tilde{E}\partial_m \tilde Q^{mnp}
\tilde Q_p^{\ st}\partial_n\epsilon_{st}\nonumber\\
&&+\tilde Q^{mnq}\tilde Q_q^{\ st}\left(\tilde{S}_{mp}^{\ \ \ p}\partial_n\epsilon_{st}
-\frac{1}{2}\tilde{S}_{mn}^{\ \ \ p}\partial_p\epsilon_{st}\right).
\end{eqnarray}
This leads to the identity
\begin{eqnarray}\label{B24}
&&\int d^dx\tilde{E}\tilde Q^{mnq}\tilde Q_q^{\ st}
(\tilde{S}_{nm}^{\ \ \ p}
\partial_p\epsilon_{st}-2\tilde{S}_{np}^{\ \ \ p}\partial_m\epsilon_{st})\nonumber\\
=2&&\int d^dx\tilde{E}\partial_n(\tilde Q^{mnp}\tilde Q_p^{\ st})\partial_m\epsilon_{st}.
\end{eqnarray}

In the difference of the invariants $I_E-I_{\Omega K}$ the inhomogeneous
piece $\sim\tilde{Q}^{mnp}\partial_m\epsilon_{np}$ cancels.
However, for this difference the
inhomogeneous piece in the local Lorentz transformation involves new fermion bilinears with
derivatives. One may try to cancel this piece by adding further invariants with higher
powers of $\tilde{S}$ and $\tilde{Q}$. It is not obvious, however, if the system can be
closed such that one could finally find a local invariant. 

Recently, a first construction leading to local invariants has been identified \cite{CWST}. It involves a maximal number of spinor fields $\psi$ without derivatives, multiplied with precisely $d$ factors $\partial_\mu \psi$. Since the inhomogeneous part of the local Lorentz-transformation of $\partial_\mu\psi$ leads to one more factor of $\psi$ (multiplied with $\partial\epsilon$) it must vanish due to the Grassmann properties of $\psi$. This construction requires that one can form a Lorentz-singlet out of the $d$ factors $\partial_\mu\psi$. We have already seen in sect. \ref{spinoraction} that this is not possible for Majorana-Weyl spinors in $d=2~mod~8$ whereas in \cite{CWST} the viability of this construction was demonstrated for $d=8~mod~8$.


\begin{thebibliography}{}
\bibitem{Ak}K. Akama, Prog. Theor. Phys. {\bf 60} (1978) 1900
\bibitem{Av}D. Amati, G. Veneziano, Phys. Lett. {\bf 105B} (1981) 358
\bibitem{DS}G. Denardo, E. Spallucci, Class. Quantum Grav. 4, 89 (1987)
\bibitem{HW1}A. Hebecker, C. Wetterich, Phys. Lett. {\bf B574} (2003) 269
\bibitem{SG1}C. Wetterich, hep-th/0307145, to appear in Phys. Rev. {\bf D}
\bibitem{CWST}C. Wetterich, hep-th/0405223
\bibitem{GenG}C. Wetterich, Nucl. Phys. {\bf B397}, 299 (1993)
\bibitem{CWGG}C. Wetterich, Nucl. Phys. {\bf B242} (1984) 473
\bibitem{CWMS}C. Wetterich, Nucl. Phys. {\bf B211} (1983) 177
\bibitem{SW}J. Schwindt, C. Wetterich, Phys. Lett. {\bf B578} (2004) 409
\bibitem{CWF} C. Wetterich, Nucl. Phys. {\bf B223} (1983) 109;\\
E. Witten, in Proc. of the 1983 Shelter Island Conference II (MIT Press,
Cambridge, Mass., 1984)
\bibitem{CWB}C. Wetterich, Nucl. Phys. {\bf B255}, 480 (1985)
\bibitem{RS}V. Rubakov, M. Shaposhnikov, Phys. Lett. {\bf B125}, 139 (1983)
\bibitem{CWFA}C. Wetterich, Nucl. Phys. {\bf B244} (1983) 359; {\bf B261} (1985) 461
\bibitem{SDE}F.~J.~Dyson, Phys. Rev. {\bf 75} (1949) 1736\\
J.~.S.~Schwinger, Proc. Nat. Acad. Sci. 37 (1951) 452
\bibitem{AW}L. Alvarez-Gaum\'e, E. Witten, Nucl. Phys. {\bf B234} (1984) 269
\bibitem{Tor}V. Gusynin, E. Gorbar, V. Romankov, Nucl. Phys. {\bf B362} (1991) 449
\bibitem{BEA}C. Wetterich, cond-mat/0208361
\bibitem{JW}J. Jaeckel, C. Wetterich, Nucl. Phys. {\bf A733} (2004) 113
\end{thebibliography}
\end{document}